\documentclass[%
 reprint,
 amsmath,amssymb,
 aps,pre,
showkeys,
floatfix,
]{revtex4-2}

\usepackage[hidelinks,bookmarks=false]{hyperref}% add hypertext capabilities
\usepackage{graphicx}% Include figure files
\usepackage[caption=false]{subfig}
\usepackage{dcolumn}% Align table columns on decimal point
\usepackage{bm}% bold math
\usepackage[thinlines]{easytable}
\usepackage{bm}
\usepackage{amsmath}
\usepackage{tikz}
\usepackage{ amssymb }
\usepackage{ dsfont }
\usepackage{comment}
\usepackage[english]{babel}

\newcommand{\matr}[1]{\mathbf{#1}}

\begin{document}
%\preprint{APS/123-QED}

\title{Community detectability and structural balance dynamics in signed networks}%

\author{Megan Morrison}
 \email{mmtree@uw.edu}
\affiliation{%
 Department of Applied Mathematics, University of Washington, Washington 98115, USA}%
\author{Michael Gabbay}%
 \email{gabbay@uw.edu}
\affiliation{%
 Applied Physics Laboratory, University of Washington, Washington 98115, USA}%

\date{\today}% It is always \today, today,

\begin{abstract}
We investigate signed networks with community structure with respect to their spectrum and their evolution under a dynamical model of structural balance, a prominent theory of signed social networks. The spectrum of the adjacency matrix generated by a stochastic block model with two equal size communities shows detectability transitions in which the community structure becomes manifest when its signal eigenvalue appears outside the main spectral band. The spectrum also exhibits ``sociality'' transitions involving the homogeneous structure representing the average tie value. We derive expressions for the eigenvalues associated with the community and homogeneous structure as well as the transition boundaries, all in good agreement with numerical results. Using the stochastically-generated networks as initial conditions for a simple model of structural balance dynamics yields three outcome regimes: two hostile factions that correspond with the initial communities, two hostile factions uncorrelated with those communities, and a single harmonious faction of all nodes. The detectability transition predicts the boundary between the assortative and mixed two-faction states and the sociality transition predicts that between the mixed and harmonious states. Our results may yield insight into the dynamics of cooperation and conflict among actors with distinct social identities.
\end{abstract}


%Use showkeys class option if keyword
\maketitle

\section{\label{sec:intro}Introduction}

Most research in network science has focused on networks that allow only positive ties. In signed networks, however, ties can take on negative values as well. In social systems, positive ties signify friendly or cooperative relationships between the individual or collective actors represented by the nodes whereas negative ties signify hostile or conflictual relationships between nodes. As examples, signed social networks have been used to represent interpersonal sentiments among students \cite{kirkley_balance_2019}, supportive or critical references among opinion makers \cite{BruTraUit2012}, relationships in online social networks \cite{facchetti_computing_2011}, and alliances and military clashes among nations \cite{Maozetal2007,DorMrv2015}.

In this paper, we address community structure in signed networks and its implications for dynamics governed by structural balance, a theory commonly invoked in treatments of signed networks in social systems. In unsigned networks, community structure refers to the presence of clusters within networks characterized by relatively dense intra-cluster ties and sparse inter-cluster ties. A rich set of techniques have been developed to detect communities in unsigned networks \cite{ForHri2016}. Of particular relevance here, spectral analysis has proven to be a highly valuable tool for probing community structure \cite{Newman2006,ChaGirOtt2009}. For signed networks, the notion of community can be extended to accommodate negative ties by reversing the criteria for positive ties --- there should be relatively sparse negative ties within communities and denser ties between them. At the present, however, the literature on community detection in signed networks is itself rather sparse in comparison with unsigned networks \citep[e.g.,][]{traag_community_2009,kunegis_spectral_2010,BruTraUit2012,EsmJal2015}.

An important phenomenon of community structure in unsigned networks is that of community detectability \cite{Decelleetal2011,nadakuditi_graph_2012,zhang_spectra_2014,Ghaetal2016,Wiletal2019}. Here, community structure can be present --- in the sense that the tie generating probabilities in a stochastic block model indeed favor ingroup over outgroup ties --- but it is too weak to typically be discerned by analysis of the generated network. For large networks, a phase transition characterizes the passage from undetectable to detectable structure.

\begin{figure}[!htbp]
\centering
\begin{tikzpicture}[scale=1]
\node[inner sep=0pt](russell) at (0,0)
 {\includegraphics[width=0.4\textwidth]{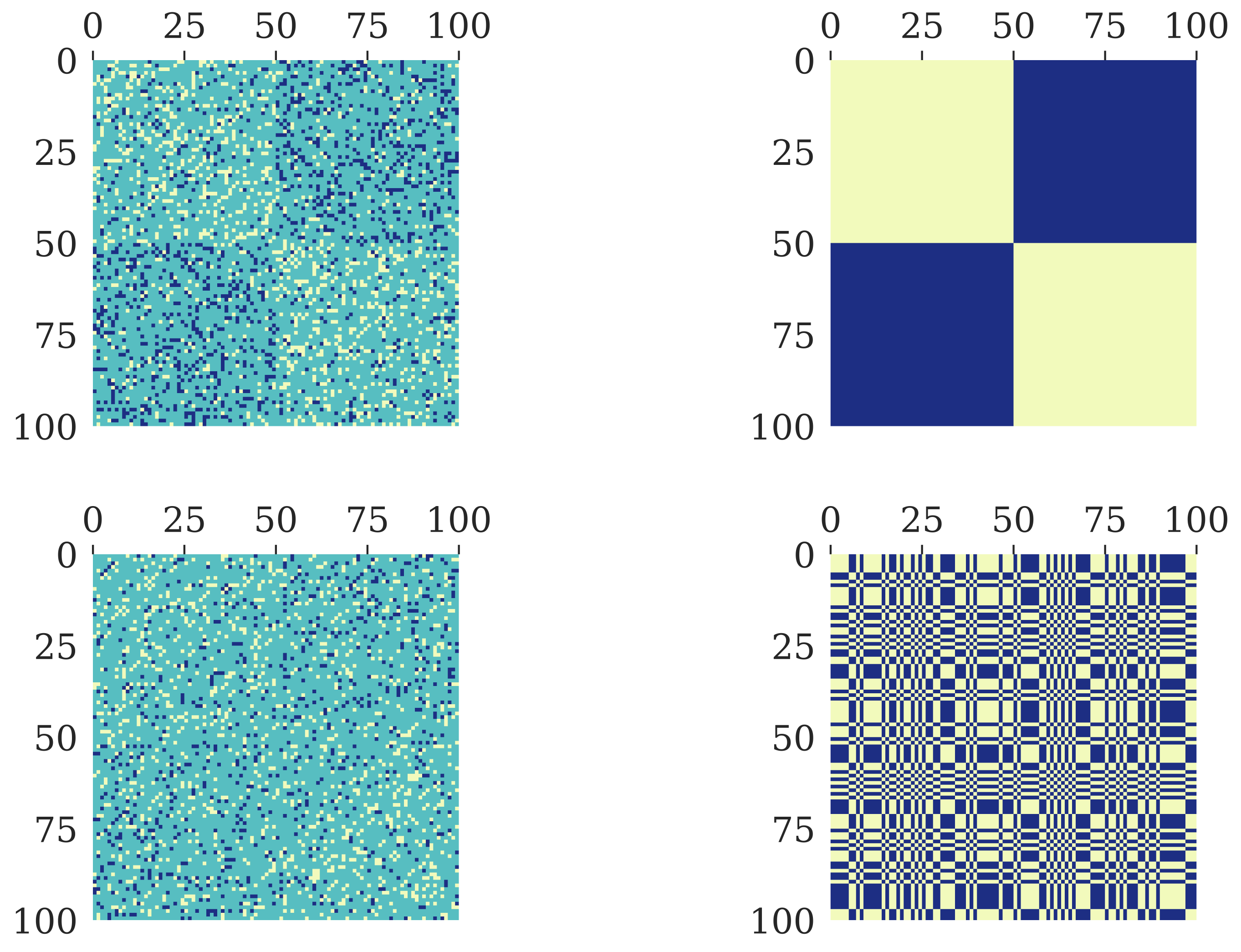}};
 \node (fig2) at (4.2,1.35)
       {\includegraphics[scale=0.08]{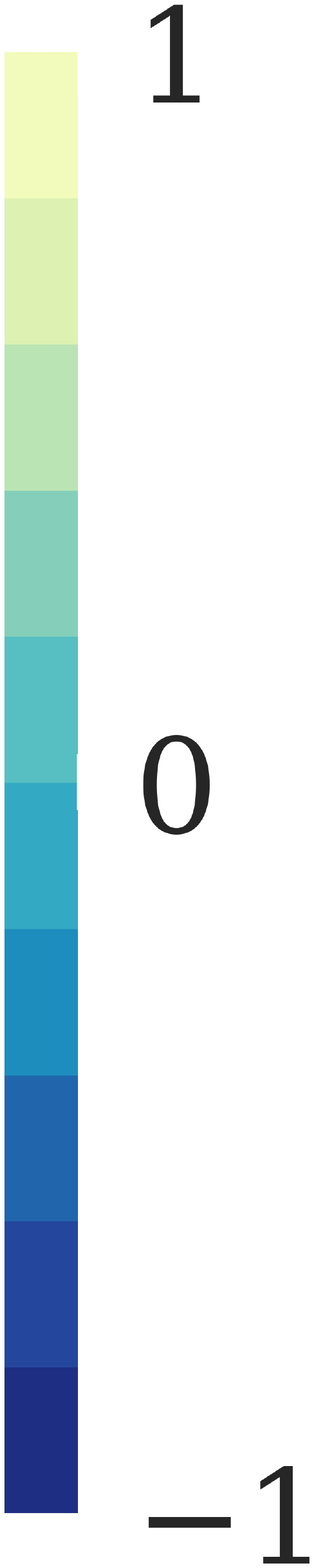}};
\node (fig2) at (-0.05,1.35)
       {\includegraphics[scale=0.08]{colorbar.pdf}};
 %\draw[fill] (-2.3,3.3) circle [radius=0.025];
 \node at (-3.75,3.0) {\small  (a)};

 \node at (-3.75,-0.25) {\small  (b)};

 \node[] at (-2,3.5) {\small Initial Matrix};
 \node[] at (2.4,3.5) {\small Final Matrix};

 \node at (-2,3.0) {\scriptsize  Type A \ \ Type B};
  \node at (2.3,3.0) {\scriptsize Type A \ \ Type B};

 \node[label=above:\rotatebox{90}{\scriptsize  Type B \ \ Type A}] at (-3.75,0) {};
 \node[label=above:\rotatebox{90}{\scriptsize  Type B \ \ Type A}] at (0.5,0) {};

 \draw[->,thick](-0.4,1.3)--(0.2,1.3);
 \draw[->,thick](-0.4,-1.5)--(0.2,-1.5);
 \end{tikzpicture}
 \caption{Evolution of networks with initial community structure under structural balance dynamics. (a) Moderate initial structuring by group identity leads to a completely connected network consisting of two factions completely sorted by identity. (b) Weak initial structure leads to two factions of mixed identities. Networks represented as adjacency matrices with $\pm 1, 0$ tie values indicated by color. Initial networks generated by stochastic block model, (\ref{eq:Ablock})--(\ref{eqn:Ao}) with parameters $d_{in} = 0.2$, $d_{out} = 0.2$, $p_{out}^+ = 0.33$, and $p_{in}^+ = 0.7$ for (a) and $d_{in} = 0.2$, $d_{out} = 0.2$, $p_{out}^+ = 0.5$, and $p_{in}^+ = 0.7$ for (b).  Final networks represent convergence of Eq.~(\ref{eq:cont time model}). }
 \label{fig:intro_mat}
\end{figure}

We show that detectability transitions also occur in signed networks. We generate our networks using a stochastic block model for two communities in an unweighted and undirected signed network (Sec.~\ref{sec:gencomm}). Examples of simulated networks with community structure that is detectable and undetectable are shown on the left in Fig.~\ref{fig:intro_mat}(a) and (b) respectively. We describe the transitions observed in the spectra of simulated networks in which outlying eigenvalues corresponding to meaningful signals merge with the main spectral band corresponding to noise (Sec.~\ref{sec:detect}). Two sets of spectral transitions are found: one corresponds to the detectability transition involving the two-community structure, while the other affects the ability to observe an overall tendency toward positive or negative tie formation, which we refer to as sociality transitions.

We analytically calculate both the key eigenvalues and the transition conditions for large networks. In the main text, we use perturbation analysis to derive expressions for the signal eigenvalues (Sec.~\ref{sec:PertCalc}), which are then used to obtain the transition conditions by their equation with the main band edge eigenvalues (Sec.~\ref{sec:trans_bound}), these edge eigenvalues being found using random matrix theory (App.~\ref{appendix1}). We also present an alternative to our perturbation treatment that derives the signal eigenvalues on the basis of random matrix theory, in keeping with previous treatments of detectability (App.~\ref{appendix2}) \cite{nadakuditi_graph_2012,zhang_spectra_2014}.

The spectral transitions have important implications for the outcomes of structural balance dynamics for networks possessing initial community structure. Structural balance theory, which postulates that triads with one or three negative edges will not endure, can be implemented as a continuous time dynamical system (Sec.~\ref{sec:StrucBalDyn}) \cite{KulGawGro2005,marvel_continuous-time_2011}. The model evolves into a fully connected network where either: (1) there are two hostile factions with only positive ties within each and only negative ties between them; or (2) all nodes are positively connected in a single harmonious faction. In either case, the final state is determined by the leading eigenvector of the initial network.

The driving role played by the leading eigenvector of the initial network in the structural balance evolution gives rise to a dynamical manifestation of the detectability transition when the leading eigenvector also carries information about community structure. For the two-faction outcome, if the leading eigenvector corresponds to the two identity types in the stochastic block model, then the  final factions will perfectly align with these identities as shown in Fig.~\ref{fig:intro_mat}(a). On the other hand, if the leading eigenvector is merely the edge of the main noise band, as occurs for weak initial structure below the detectability transition, then the composition of the final factions will not align with the identity types as seen in Fig.~\ref{fig:intro_mat}(b). An analogous transition to the single-faction outcome is generated by the sociality transition. Solutions of the structural balance model starting from networks randomly generated by the stochastic block model do indeed show sharp transitions between behavioral regimes whose boundaries agree with analytical predictions based on the detectability and sociality transitions (Sec.~\ref{sec:regimes}).

We discuss the potential implications of these results for conflict dynamics among actors with different identity types due to, for instance, ethnicity, religion, or ideology (Sec.\ref{sec:discussion}). In particular, conflicts such as civil wars may take on a binary nature. If the system starts out with weak identity-driven structure, then it will not be expected to polarize on the basis of identity. But complete identity polarization results even when initial affinities and animosities between identity types are fairly mild and even though identity itself plays no role in the micro-level conflict dynamics.

%%%%%% tikz %%%%%%%%%%%%%
%

%%%%%%%%%%%%%%%%%%%%%%%
%%%% New Section %%%%%%
%%%%%%%%%%%%%%%%%%%%%%%
\section{\label{sec:gencomm}Generating and representing community structure}

Communities in an unsigned network are characterized by relatively dense within-community ties and sparse ties between communities. Community detection algorithms seek to discover these communities given an observed network \cite{GirNew2002,newman_finding_2006,ForHri2016}. Stochastic block models, which generate random networks with community structure by setting tie probabilities within and between blocks of nodes, have been used to investigate the behavior of community detection algorithms \cite{newman_networks_2018}. In this section, we describe the stochastic block model we use to generate our signed networks, the characterization of community structure via assortativity, and decomposition of the generated networks in terms of the eigenvectors of the average adjacency matrix and a random matrix.

%
%%%%%%%%%%%%%%%%%
%% subsection %%%
%%%%%%%%%%%%%%%%%
\subsection{\label{sec:SBM}Stochastic block model}

Our construction starts with an undirected network of $N$ nodes consisting of two identity groups A and B of equal size $N/2$, where $N \gg 1$. The A group nodes are indexed from 1 to $N/2$ and the B group from $N/2+1$ to $N$. $\matr{A}$ is the signed adjacency matrix where $\matr{A}_{ij}$ is the tie value between node $i$ and node $j$, which can take on values of $\{1,-1,0\}$ with 0 signifying the absence of a tie. As the network is undirected, the adjacency matrix is symmetric, $\matr{A}_{ij} = \matr{A}_{ji}$. The probability that a tie, positive or negative, will form between any given ingroup (A with A, B with B) node pair is $d_{in}$. Similarly, the tie formation probability between outgroup (A with B) node pairs is $d_{out}$. These tie formation probabilities are equivaent to the expected ingroup and outgroup tie densities and their average yields the expected tie density for the total network, $d=(d_{in}+d_{out})/2$. Given the presence of a tie between ingroup members, the conditional probability that it is positive is $p_{in}^+$ and that it is negative is $p_{in}^-=1-p_{in}^+$. Similarly, the positive and negative tie conditional probabilities between outgroup nodes are written $p_{out}^+$ and $p_{out}^-=1-p_{out}^+$. For brevity, we refer to $p_{in}^+$ and $p_{out}^+$ as the ingroup and outgroup \emph{affinities} and $p_{in}^-$ and $p_{out}^-$ as the in and outgroup \emph{animosities}.

The adjacency matrix can be written in terms of the following block structure:
\begin{align}
    \matr{A} = \begin{bmatrix}
    \matr{A_{AA}}& \matr{A_{AB}}\\
    \matr{A_{BA}}& \matr{A_{BB}}
    \end{bmatrix}, \label{eq:Ablock}
\end{align}
where each block is a random $N/2 \times N/2$ matrix. The diagonal blocks represent AA or BB ties, whose elements are set using the following probability distribution for the ingroup random variable $A_{in}$:
\begin{align}
    \mathds{P}(A_{in}=k) &=\left\{
      \begin{array}{@{}ll@{}}
        d_{in} p_{in}^+ & ,k=1 \\
        d_{in} (1-p_{in}^+) & ,k=-1\\
        1-d_{in} & ,k=0.
      \end{array}\right. 
    \label{eqn:Ai}
\end{align}
Since $\matr{A}$ is symmetric, there are $(N/2)(N/2+1)$ independent, identically distributed ingroup ties. Note that nonzero self-ties are allowed unlike in many empirical networks. For large $N$, however, our results will still be applicable to the zero-diagonal case.

The off-diagonal blocks, corresponding to AB or BA ties, are transposes of each other resulting in $N^2/4$ \emph{i.i.d.} outgroup ties, which are drawn according to the random variable $A_{out}$:

\begin{align}
  \mathds{P}(A_{out}=k) &=\left\{
  \begin{array}{@{}ll@{}}
    d_{out} p_{out}^+ & ,k=1 \\
    d_{out} (1-p_{out}^+) & ,k=-1\\
    1-d_{out} & ,k=0.
  \end{array}\right.
  \label{eqn:Ao}
\end{align}

%%%%%%%%%%%%%%%%%%%%%%%%%%%%%%%%%%%%%%%%%%%%%%%%%%%%%%%%%
\subsection{\label{sec:assort}Assortativity}

Assortativity refers to the tendency for nodes of the same type to be more strongly connected than nodes of different types. We extend the standard definition of the assortativity coefficient for discrete node types \cite{newman_mixing_2003} to our signed network case by calculating separate coefficients for the positive and negative tie networks and then essentially differencing them. We will use the signed network assortativity coefficient to characterize the regimes of the structural balance dynamics in Sec.~\ref{sec:regimes}.

\begin{figure}[!htbp]
\centering
\begin{tikzpicture}[scale=1]
\node[inner sep=0pt](russell) at (0,0)
 {\includegraphics[width=0.3\textwidth]{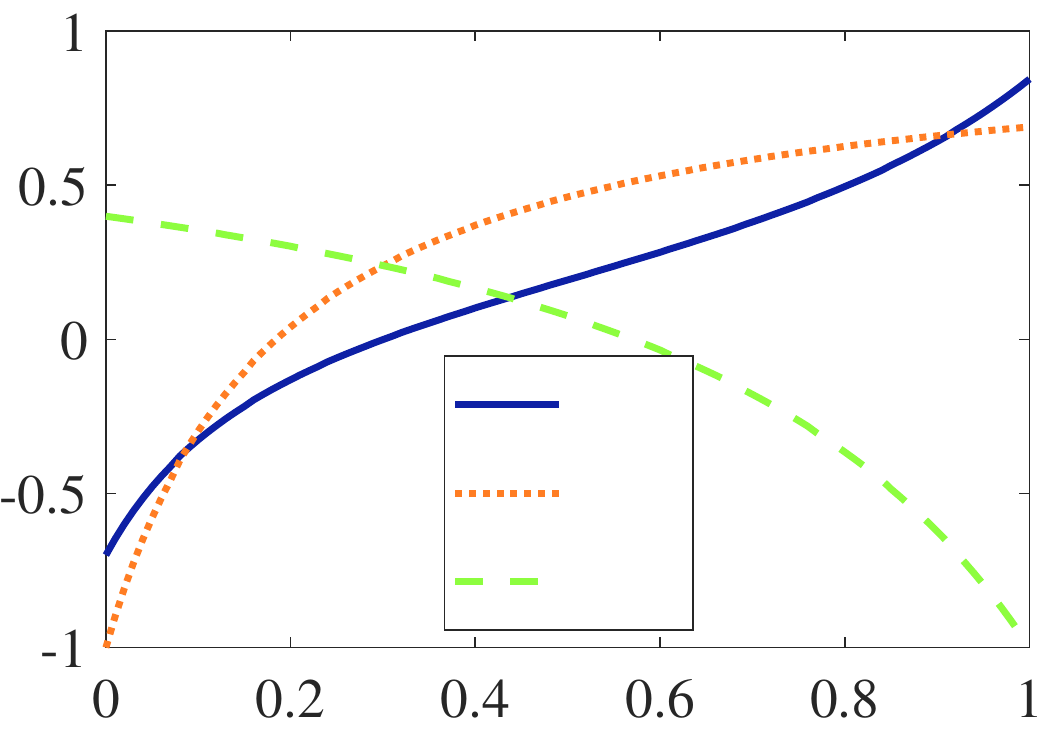}};
 \node[] at (0.4,-0.2) {\small $r$};
 \node[] at (0.5,-0.6) {\small $r^+$};
 \node[] at (0.5,-1.1) {\small $r^-$};

 \node[] at (0.2,-2.2) {$p_{in}^+$};
 \node[] at (-2.8,0.1) {$r$};
 \end{tikzpicture}
 \caption{Example of assortativity coefficients as a function of ingroup affinity. Values calculated using expected tie numbers. Parameters are $d_{in} = 0.5$, $d_{out} = 0.3$, $p_{out}^- = 0.7$, and $N=100$.}
 \label{fig:assort_ped}
\end{figure}
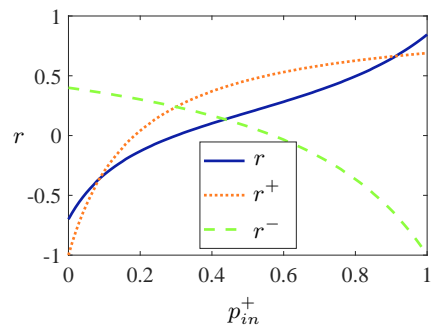

First, considering the adjacency matrix of positive ties only, we let $e^+_{ij}$ denote the fraction of all positive ties that connect a node of type $i$ to one of type $j$ where $i,j \in \{A,B\}$. The assortativity coefficient $r^+$ for the network of positive ties, whose adjacency matrix elements are 1 if $A_{ij}>0$ and zero otherwise, is then
\begin{equation}
    r^+ = \frac{\sum_i e^+_{ii} - \sum_i (a^+_i)^2}{1-\sum_i (a^+_i)^2},
    \label{eq:rpos}
\end{equation}
where $a^+_i = \sum_j e^+_{ij}$. The assortativity coefficient can range between $-1$ and 1.  A network containing only ingroup (AA or BB) positive ties with no outgroup (AB,BA) ties is completely assortative, $r^+=1$, which in the social network context implies that people only cooperate with members of the same group. A network containing only outgroup ties is completely disassortative, $r^+=-1$, implying cooperation across the two groups but not within them. We see this state in the example shown in Fig.~\ref{fig:assort_ped} in which $r^+=-1$ when the ingroup affinity $p_{in}^+=0$. As the ingroup affinity increases, $r^+$ increases but does not reach one as there are still outgroup ties due to the nonzero value of the fixed outgroup affinity.

The assortativity for the network of negative ties, $r^-$, is defined analogously to Eq.~(\ref{eq:rpos}). Whereas positive values of $r^+$ imply that ingroup relations are more friendly than outgroup relations, assortative mixing in the network of negative ties implies more hostility within groups than between them. Thus, in Fig.~\ref{fig:assort_ped} we see that $r^->0$ when there is complete ingroup animosity ($p_{in}^-=1$ corresponding to $p_{in}^+=0$) and $r^-=-1$ when there is no ingroup animosity and so negative ties within groups.

Accordingly, as we want positive values of our overall signed network assortativity coefficient $r$ to signify that ingroup interactions tend to be more amicable than outgroup ones, we average $r^+$ and $-r^-$, yielding
\begin{equation}
    r = \frac{r^+ - r^-}{2},
    \label{eq:assort}
\end{equation}
which can take on values between $-1$ and 1. Figure~\ref{fig:assort_ped} shows that $r$ is negative for low ingroup affinity and positive for high ingroup affinity.

%%%%% tikz %%%%%%%%%%%%%

%
%
%
%%%%%%%%%%%%%%%%%%
%% new section %%%
%%%%%%%%%%%%%%%%%%
\subsection{Adjacency matrix decomposition} \label{sec:decomp}

This section presents a decomposition of the adjacency matrices generated by the stochastic block model into a signal component that results from the expected tie values generated by the ingroup and outgroup random variables, $A_{in}$ and $A_{out}$, and a noise component due to random deviations from the expected values. This decomposition will form the starting point for our calculation of the network eigenvalues in the following section.

We write $\matr{A}$ as the sum of the average matrix $\langle \matr{A} \rangle$ and a random deviation matrix $\matr{X}$:
\begin{align}
    \matr{A} &= \langle \matr{A} \rangle + \matr{X}. \label{eqn:A_sum}
\end{align}
Given the block structure of Eq.~(\ref{eq:Ablock}), $\langle \matr{A} \rangle$ can be written as
\begin{align}
    \langle \matr{A} \rangle = \begin{bmatrix}
    \langle \matr{A}_{AA} \rangle & \langle \matr{A}_{AB} \rangle\\
    \langle \matr{A}_{BA} \rangle & \langle \matr{A}_{BB} \rangle
    \end{bmatrix}, \label{eq:avAblock}
\end{align}
where each element of $\langle \matr{A}_{AA} \rangle$ and $\langle \matr{A}_{BB} \rangle$ is equal to $\langle A_{in} \rangle$ and each element of $\langle \matr{A}_{AB} \rangle$ and $\langle \matr{A}_{BA} \rangle$ is equal to $\langle A_{out} \rangle$. From Eqs.~(\ref{eqn:Ai}) and (\ref{eqn:Ao}), we have
\begin{eqnarray}
\langle A_{in} \rangle & = & d_{in}(2p_{in}^+-1) \label{eq:avAi} \\
\langle A_{out} \rangle & = & d_{out}(2p_{out}^+-1). \label{eq:avAo}
\end{eqnarray}

We define a couple of useful linear combinations of the in and outgroup expected tie values. We denote by $\mu$ the average over all the elements in  $\langle \matr{A} \rangle$,
\begin{align}
\mu = \frac{\langle A_{in} \rangle + \langle A_{out} \rangle}{2}, \label{eq:mu}
\end{align}
and we denote by $\nu$ the half-difference between the in and outgroup expected tie values,
\begin{align}
\nu = \frac{\langle A_{in} \rangle - \langle A_{out} \rangle}{2}. \label{eq:nu}
\end{align}
Both $\mu$ and $\nu$ range from $-1$ to 1. Noting that $\langle A_{in} \rangle = \mu+\nu$ and $\langle A_{out} \rangle = \mu-\nu$, these expressions allow us to express $\langle \matr{A} \rangle$ as a sum of two outer products,
\begin{align}
\label{eq:Aavg_clean}
    \langle \matr{A} \rangle &=  \mu N\matr{u}_H \matr{u}_H^T +  \nu N \matr{u}_C \matr{u}_C^T,
\end{align}
where $\matr{u}_H = \frac{1}{\sqrt{N}}[1, 1, ..., 1]^T$ and $\matr{u}_C = \frac{1}{\sqrt{N}}[1,..., 1, -1,..., -1]^T$ are orthonormal $N$-dimensional vectors. In fact, $\matr{u}_H$ and $\matr{u}_C$ are readily seen to be the two eigenvectors of the rank 2 matrix $\langle \matr{A} \rangle$ with respective eigenvalues $\mu N$ and $\nu N$:
\begin{align}
    \langle \matr{A} \rangle \matr{u}_H &= \mu N \matr{u}_H  \label{eq:uHEigEq} \\
    \langle \matr{A} \rangle \matr{u}_C &= \nu N \matr{u}_C. \label{eq:uFEigEq}
\end{align}

The term containing $\matr{u}_H$ in Eq.~(\ref{eq:Aavg_clean}) generates a homogeneous $N \times N$ matrix whose elements are all equal to $\mu$, the global average tie value. Hence, we refer to $\matr{u}_H$ as the \emph{homogeneous} eigenvector. The term containing $\matr{u}_C$ generates a matrix whose diagonal block elements are all equal to $\nu$ and whose off-diagonal block elements are $-\nu$ and so corresponds to the structure of ingroup and outgroup tie differences. Accordingly, $u_C$ generates the community structure and we refer to it as the \emph{contrast} eigenvector. The homogeneous and contrast eigenvectors are \emph{signal} eigenvectors whose ability to be distinguished from the noise generated by $\matr{X}$ has important implications for community detectability and structural balance dynamics. From this perspective, $\mu$ and $\nu$ can be regarded as natural parameters for the signal structure in the network and could be used in place of two of the parameters in the stochastic block model, for instance, the ingroup and outgroup affinities. Doing so is less intuitive from a simulation viewpoint, however.

The noise matrix $\matr{X}$ is a symmetric matrix that can be written in the block form,
\begin{align}
     \matr{X} &= \begin{bmatrix}
    \matr{X}_{AA}& \matr{X}_{AB}\\
    \matr{X}_{BA} & \matr{X}_{BB}
    \end{bmatrix}. \label{eq:Xblock}
\end{align}
Since $\matr{X} = \matr{A}-\langle \matr{A} \rangle$, the elements of the ingroup blocks $\matr{X}_{AA}$ and $\matr{X}_{BB}$ can assume values in $\{1-\langle A_{in} \rangle, -1 -\langle A_{in} \rangle, -\langle A_{in} \rangle \}$ that are distributed according to the random variable $X_{in}$,
\begin{align}
\label{eqn:Xi}
    \mathds{P}(X_{in}=k) &=\left\{
      \begin{array}{@{}ll@{}}
        d_{in} p_{in}^+ & ,k=1-\langle A_{in} \rangle \\
        d_{in} (1-p_{in}^+) & ,k=-1 -\langle A_{in} \rangle\\
        1-d_{in} & ,k=-\langle A_{in} \rangle .
      \end{array}\right.
\end{align}
Likewise, the entries of the outgroup blocks $\matr{X}_{AB}=\matr{X}^T_{BA}$ are distributed like $X_{out}$,
\begin{align}
      \label{eqn:Xo}
     \mathds{P}(X_{out}=k) &=\left\{
  \begin{array}{@{}ll@{}}
    d_{out} p_{out}^+ & ,k=1-\langle A_{out} \rangle \\
    d_{out} (1-p_{out}^+) & ,k=-1-\langle A_{out} \rangle \\
    1-d_{out} & ,k=-\langle A_{out} \rangle .
  \end{array}\right.
\end{align}
All the elements of $\matr{X}$ have zero mean as $\langle X_{in} \rangle = \langle X_{out} \rangle = 0$. The variances of $X_{in}$ and $X_{out}$ are given by $\sigma^2_{in}=\langle X_{in}^2 \rangle$ and $\sigma^2_{out} =\langle X_{out}^2 \rangle$, which are written in terms of the stochastic block model parameters as
\begin{align}
&\sigma^2_{in}  = d_{in} - d_{in}^2(2p_{in}^+-1)^2 \label{eq:varXi} \\
&\sigma^2_{out} = d_{out} - d_{out}^2(2p_{out}^+-1)^2. \label{eq:varXo}
\end{align}
These variances will appear as their average,
\begin{equation}
\sigma^2=\frac{\sigma^2_{in}+\sigma^2_{out}}{2},
\label{eq:sigma}
\end{equation}
in the noise-induced correction to the signal eigenvalues calculated below. The average variance can also be related to the parameters $\mu$ and $\nu$ as follows,
\begin{align}
\sigma^2 &= \frac{d_{in} + d_{out}}{2} - \frac{d_{in}^2(2p_{in}^+-1)^2 + d_{out}^2(2p_{out}^+-1)^2}{2} \label{eq:sigma2pars} \\
         &= \frac{d_{in} + d_{out}}{2} - \frac{1}{2}(\langle A_{in} \rangle^2 + \langle A_{in} \rangle^2) \\
         &= \frac{d_{in} + d_{out}}{2} - \mu^2 - \nu^2, \label{eq:sigma2munu}
\end{align}
where we have used Eqs.~(\ref{eq:avAi}) and (\ref{eq:avAo}) in the second line and $\langle A_{in} \rangle^2 + \langle A_{out} \rangle^2 = (\mu + \nu)^2 + (\mu - \nu)^2  = 2(\mu^2 + \nu^2)$ in the third.

%%%%%%%%%%%%%%%%%%
%% new section %%%
%%%%%%%%%%%%%%%%%%
\section{Detectability of signal eigenvalues} \label{sec:detect}
Spectral analysis has been used to address the number and detectability of communities in unsigned networks by considering the leading eigenvalues that reside outside the (approximately) continuous main spectral band due to its generation as a random graph \cite{ChaGirOtt2009,nadakuditi_graph_2012, zhang_spectra_2014, sarkar_spectral_2018}. For undirected networks, the adjacency matrix is symmetric and hence has a real spectrum. The number of detectable communities is equivalent to the number of positive eigenvalues that lie beyond the main spectral band. \citet{nadakuditi_graph_2012} showed the existence of, and analytically calculated, a detectability transition in which the community structure, as generated by a stochastic block model with two communities, while still present becomes no longer detectable. This transition occurs once the second eigenvalue of the adjacency matrix, which carries the community structure information, merges with the main spectral band. Using random matrix theory, the authors derived expressions for both leading eigenvalues and the edge of the spectral band, thereby enabling the analytical determination of the transition dependence.

\begin{figure}[!htbp]
\centering
\begin{tikzpicture}[scale=1]
\node[inner sep=0pt](russell) at (0,0)
 {\includegraphics[width=0.9\linewidth]{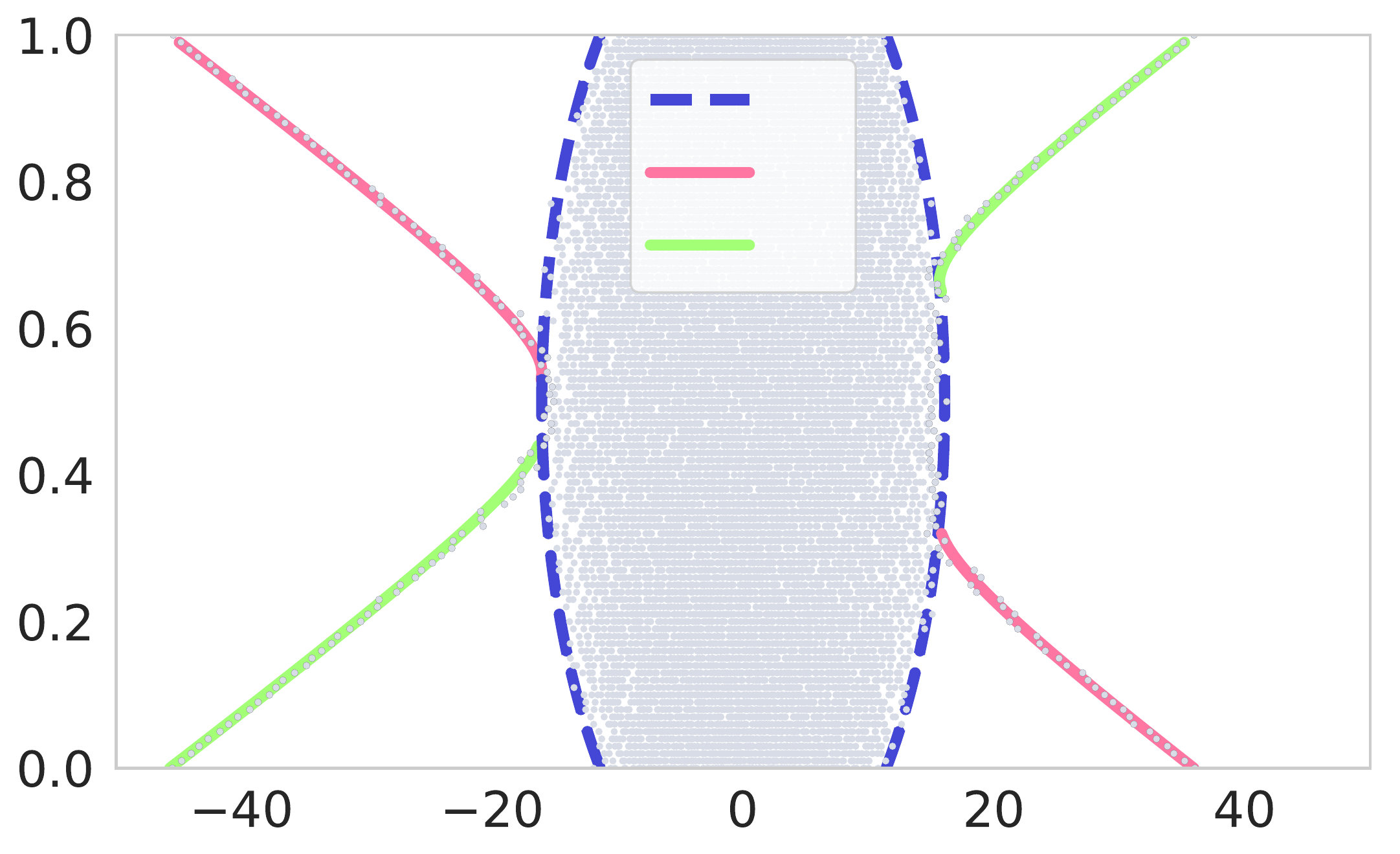}};
 \node at (-4.2,2.2) {(a)};
 \node at (0.6,1.8) {$\gamma$};
 \node at (0.6,1.4) {$\lambda_H$};
 \node at (0.6,1.0) {$\lambda_C$};
 \node at (-4.2,0.1) {$p_{out}^-$};
 \node at (0.25,-2.6) {$\lambda(\matr{A})$};

 \node at (2.8,1.0) {\footnotesize assortative};
 \node at (2.8,0.7) {\footnotesize transition};
 \draw[->,thick](2,0.83)--(1.5,0.83);

 \node at (2.8,-0.4) {\footnotesize prosocial};
 \node at (2.8,-0.7) {\footnotesize transition};
 \draw[->,thick](2,-0.51)--(1.5,-0.51);

  \node at (-2.3,0.67) {\footnotesize antisocial};
 \node at (-2.3,0.37) {\footnotesize transition};
 \draw[->,thick](-1.35,0.35)--(-1.0,0.35);

 \node at (-2.3,-0.05) {\footnotesize disasortative};
 \node at (-2.3,-0.35) {\footnotesize transition};
  \draw[->,thick](-1.35,-0.06)--(-1.0,-0.06);
\end{tikzpicture}
\begin{tikzpicture}[scale=1]
\node[inner sep=0pt](russell) at (0,0)
 {\includegraphics[width=0.9\linewidth]{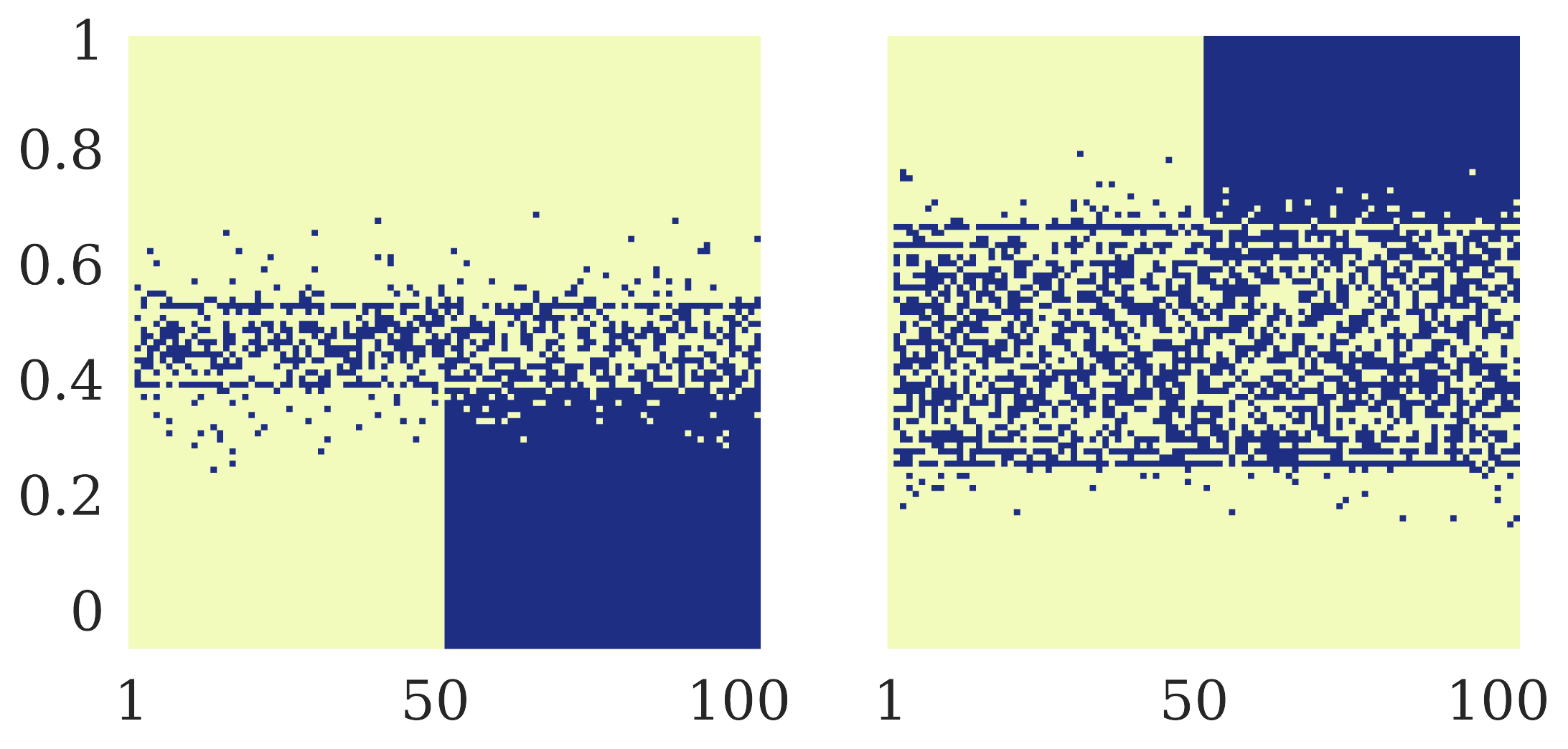}};
 \node at (-4.2,1.8) {(b)};
 \node at (-1.7,-2) {$\matr{u}_N$};
 \node at (2.05,-2) {$\matr{u}_1$};
 \node at (-4.2,0.1) {$p_{out}^-$};
\end{tikzpicture}
\caption{(a) Eigenvalue spectrum of $\matr{A}$ as a function of outgroup animosity $p_{out}^-$. The other parameters remain constant: $p_{in}^+ = 0.4$, $d_{in} = 0.5$, $d_{out} = 0.8$. (b) Signs (yellow positive, blue negative) of the components of the last, $\matr{u}_N$, and first, $\matr{u}_1$, eigenvectors  as function of $p_{out}^-$. The theoretical curves for $\lambda_C$ (solid green), $\lambda_H$ (solid pink), and $\gamma$ (dashed blue) are calculated from Eqs.~(\ref{eq:lamFpert}), (\ref{eq:lamHpert}), and (\ref{eq:gamma2}) respectively. The diagonal elements $A_{ii}=0$ as is more usual in social networks than the nonzero values allowed in the stochastic block model.
}
\label{fig:zH_zF_spectra}
\end{figure}
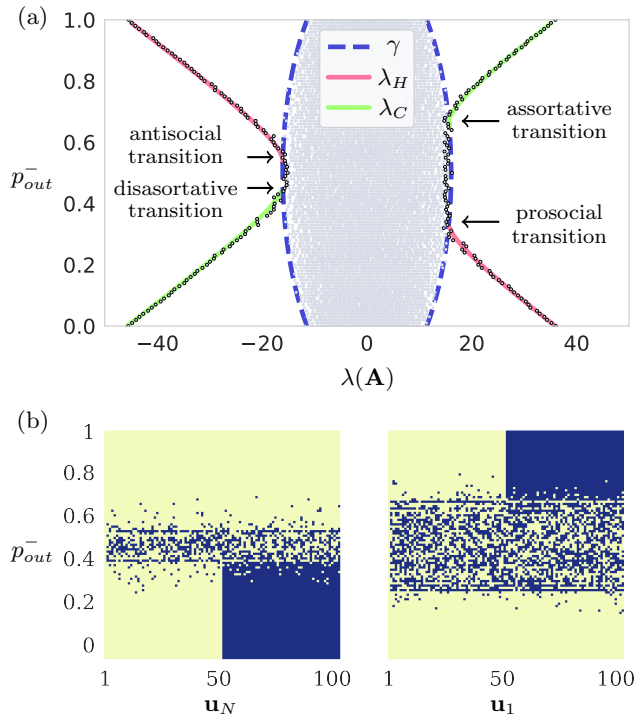

Similar to unsigned networks, Fig.~\ref{fig:zH_zF_spectra} illustrates that the spectra of our signed networks consist of a continuous band of eigenvalues originating from $\matr{X}$, the variability or noise in the system, and signal eigenvalues originating from $\langle \matr{A} \rangle$, the structure in the system. The successive horizontal slices in Fig.~\ref{fig:zH_zF_spectra}(a) correspond to the eigenvalues of single instances of $\matr{A}$ generated by the stochastic block model as the outgroup animosity is increased. There are four points at which the outlying eigenvalues merge with the main band. Considering first the upper right of the plot, the largest eigenvalue, $\lambda_1$, is observed to detach from the main band for $p^-_{out}$ greater than about 0.7. The right plot in Fig.~\ref{fig:zH_zF_spectra}(b), which depicts the signs of components of the first eigenvector $\matr{u}_1$, shows that $\matr{u}_1$ displays a two block structure for high $p^-_{out}$. Consequently, in this regime, the leading eigenvector corresponds to a perturbed version of the contrast eigenvector, $\matr{u}_C$, of $\langle \matr{A} \rangle$. The point at which $\lambda_1$ emerges from the main band is then identified with the community detectability transition. Indeed, below this point $\matr{u}_1$ loses its block structure and rapidly takes on the appearance of random noise. We observe that in contrast to the analogous case in unsigned networks where the detectability transition involves the second eigenvalue of the adjacency matrix, here the two communities are no longer discernible when the leading eigenvalue merges with the main spectral band. We refer to this transition as the \emph{assortative} transition because the communities are preferentially grouped by identity type. With an eye toward its dynamical significance, the two final factions produced by the structural balance dynamics are polarized by identity type after $\lambda_1$ emerges from the band.

The leading eigenvalue is observed to undergo another transition at $p^-_{out} \approx 0.3$. For lower outgroup animosities, $\matr{u}_1$ takes on a single block structure and so can be taken to result from a perturbation of the homogeneous eigenvector $\matr{u}_H$. This homogeneous structure disappears from $\matr{u}_1$ for $p^-_{out}$ values above the transition. As the homogeneous eigenvalue carries information about the average tie value over all nodes, its emergence from the noise band can be considered a \emph{sociality} transition. In particular, we refer to transitions that occur on the positive side of the noise band as \emph{prosocial} transitions, in which a pattern of overall positive ties between nodes becomes apparent. The prosocial transition induces a transition in the structural balance dynamics from the two-faction equilibrium (not sorted by identity) to a single harmonious faction consisting of all nodes.

The lower left section of Fig.~\ref{fig:zH_zF_spectra}(a) shows the intersection of the last and least eigenvalue, $\lambda_N$, with the noise band at $p^-_{out} \approx 0.4$. For outgroup animosity values beneath this intersection, the last eigenvector $\matr{u}_N$ displays a two block structure as seen on the left plot of Fig.~\ref{fig:zH_zF_spectra}(b). However, although these blocks align with the A and B identity groups, the warmer outgroup than ingroup relations implies that the blocks are really disassortative ``anti''-communities rather than assortative communities (prominent negative eigenvalues are also associated with disassortativity in unsigned networks \cite{newman_finding_2006}). Since this \emph{disassortative} transition involves the least eigenvector, it has no significance with respect to the outcomes of the structural balance dynamics.

Finally, the other transition involving $\lambda_N$, seen in the upper left of Fig.~\ref{fig:zH_zF_spectra}(a), represents the emergence of the homogeneous eigenvalue and the corresponding single block structure from the noise band. It is a sociality transition and, in particular, an \emph{antisocial} transition as it occurs on the negative side of the noise band signifying a conflictual relationship among nodes on average. The antisocial transition has no dynamical significance with respect to the structural balance dynamics.
%
%%%%%%%%% tikz %%%%%%%%%

\section{Calculation of signal eigenvalues} \label{sec:PertCalc}
\par
In this section, we will derive formulas for the signal eigenvalues as a function of our stochastic block model parameters. We employ a perturbation treatment here but present an alternative derivation employing random matrix theory and complex analysis in App.~\ref{appendix2}. Equation~(\ref{eqn:A_sum}), which expresses the adjacency matrix $\matr{A}$ as the sum of its expected value $\langle \matr{A} \rangle$ and a matrix of random deviations $\matr{X}$, will form the starting point of our analysis. We consider $\langle \matr{A} \rangle$ as a given deterministic matrix with homogeneous and contrast eigenvalues and eigenvectors, Eqs.~(\ref{eq:uHEigEq}) and (\ref{eq:uFEigEq}), that is subject to a perturbation from the independent noise matrix $\matr{X}$, which induces shifts to the signal eigenvalues and eigenvectors. A perturbation expansion to second order and the statistics of the noise matrix then yield the corrections to the signal eigenvalues. We note that in treating $\matr{X}$ as an independent perturbation to $\langle \matr{A} \rangle$, we temporarily suspend their linkage via the tie formation probabilities in the stochastic block model.
\subsection{Perturbation Expansion Setup} \label{sec:pertsetup}
We show the perturbation calculation for the case of the contrast eigenvalue. The homogeneous eigenvalue can be obtained in precisely analogous fashion. The eigenvalue equation is
\begin{equation}
(\langle \matr{A} \rangle + \matr{X}) \matr{v}_C = \lambda_C \matr{v}_C. \label{eq:eigdef}
\end{equation}
We write the perturbed eigenvector and eigenvalue up to second order as
\begin{eqnarray}
    \matr{v}_C & = & \matr{u}_C + \matr{u}^{(1)} + \matr{u}^{(2)} \label{eq:vF} \\
    \lambda_C & = & \nu N + \lambda^{(1)} + \lambda^{(2)}, \label{eq:lambdaF}
\end{eqnarray}
where $\lambda^{(1)},\lambda^{(2)}, \matr{u}^{(1)},\matr{u}^{(2)}$ are the first and second order perturbations. As the unperturbed eigenvalue is $\mathcal{O}(\nu N)$ we divide Eq.~(\ref{eq:eigdef}) by $N$ so that the zeroth order equation is $\mathcal{O}(1)$. Then, using Eqs.~(\ref{eq:vF}) and (\ref{eq:lambdaF}), the eigenvalue equation becomes
\begin{equation}
\label{eq:eigval_pert_normed}
\begin{split}
    &\biggl(\frac{\langle \matr{A} \rangle}{N} + \frac{\matr{X}}{N}\biggl)(\matr{u}_C + \matr{u}^{(1)} + \matr{u}^{(2)})\\
    = &\biggl(\nu +
    \frac{\lambda^{(1)}}{N} + \frac{\lambda^{(2)}}{N}\biggl)(\matr{u}_C + \matr{u}^{(1)} + \matr{u}^{(2)}).
\end{split}
\end{equation}

Before embarking upon our perturbation analysis, we specify the appropriate expansion parameter. To do so, we determine the orders of the $\langle \matr{A} \rangle$ and $\matr{X}$ matrices by evaluating their 2-norms. The 2-norm of our matrix is equivalent to its largest eigenvalue. Consequently, for the unperturbed matrix, $||\langle \matr{A} \rangle||_2$ is given by the larger of $N|\nu|$ or $N|\mu|$. Taking $\nu,\mu$ to be $\mathcal{O}(1)$ therefore implies that $\langle \matr{A} \rangle$ is $\mathcal{O}(N)$. In Appendix~\ref{appendix1}, using Wigner's semicircle law for the spectral density of a random matrix as well as matrix bounds, we determine that $||\matr{X}||_2$ is $\mathcal{O}(\sigma \sqrt{N})$. The ratio of the orders of $\matr{X}$ to $\langle \matr{A} \rangle$ is $\mathcal{O}(\sigma/\sqrt{N})$, and so successive orders in the perturbation series must diminish by a factor of $\sigma/\sqrt{N}$, which therefore serves as our expansion parameter. At a given $N$, the perturbation can be made arbitrarily small by letting $\sigma$ go to zero. But in the large $N$ regime, we need not constrain $\sigma$ to be small.

Separating Eq.~(\ref{eq:eigval_pert_normed}) out by expansion orders yields
 \bigskip
 \par
\textit{$\mathcal{O}(1)$}
\begin{align}
 \frac{\langle \matr{A} \rangle}{N} \matr{u}_C = \nu \matr{u}_C
\end{align}
\bigskip
\par
\textit{$\mathcal{O}(\sigma/\sqrt{N})$}
\begin{align}
\label{eq:1st_order}
    \frac{\langle \matr{A} \rangle}{N} \matr{u}^{(1)} + \frac{\matr{X}}{N} \matr{u}_C = \nu \matr{u}^{(1)} + \frac{\lambda^{(1)}}{N} \matr{u}_C
\end{align}
\bigskip
\par
\textit{$\mathcal{O}(\sigma^2/N)$}
\begin{align}
\label{eq:2nd_order}
    \frac{\langle \matr{A} \rangle}{N} \matr{u}^{(2)} + \frac{\matr{X}}{N} \matr{u}^{(1)}
    = \nu \matr{u}^{(2)} + \frac{\lambda^{(1)}}{N} \matr{u}^{(1)} + \frac{\lambda^{(2)}}{N} \matr{u}_C.
\end{align}

\subsection{First Order Treatment} \label{sec:firstorder}

To find the first order eigenvalue perturbation, $\lambda^{(1)}$, we multiply both sides of Eq.~(\ref{eq:1st_order}) by $N \matr{u}_C^T$. Then using Eq.~(\ref{eq:Aavg_clean}) and the orthonormality of $\matr{u}_C$ and $\matr{u}_H$ gives
\begin{align}
    \nu N \matr{u}_C^T \matr{u}^{(1)} + \matr{u}_C^T \matr{X} \matr{u}_C &= \nu N \matr{u}_C^T \matr{u}^{(1)} + \lambda^{(1)}.
\end{align}
Solving for $\lambda^{(1)}$ yields
\begin{align}
    \lambda^{(1)} &= \matr{u}_C^T \matr{X} \matr{u}_C \label{eq:xux}\\
   &= \frac{1}{N} \left\{\sum_{i,j = 1}^{\frac{N}{2}}X_{ij}
    + \sum_{i,j = \frac{N}{2}+1}^{N}X_{ij}
    - 2 \sum_{i = 1}^{\frac{N}{2}}\sum_{j=\frac{N}{2}+1}^{N}X_{ij}\right\}. \nonumber 
\end{align}
The first two terms in the braces above sum ties in the ingroup blocks AA and BB respectively, each tie distributed as $X_{in}$, and the third term corresponds to the AB and BA outgroup ties, distributed as $X_{out}$. As $\lambda^{(1)}$ is equal to the sum of zero-mean random ingroup and outgroup variables, its mean therefore vanishes,
\begin{equation}
\langle \lambda^{(1)} \rangle = 0.
\label{eq:lamda1mean}
\end{equation}
Turning to the variance, each element within the outgroup sum has variance $\sigma^2_{out}$, which becomes $4\sigma^2_{out}/N^2$ when the $2/N$ prefactor is included. The contribution to the variance from the $N^2/4$ outgroup variables is therefore $\sigma^2_{out}$. Similarly, for the ingroup sums, the symmetry of $X$ implies that, neglecting the diagonal, there are approximately a total of $N^2/4$ independent variables each with variance $4\sigma^2_{in}/N^2$  so that the ingroup variance contribution is $\sigma^2_{in}$. Accordingly, the variance of $\lambda^{(1)}$ is
\begin{equation}
\mathrm{Var}(\lambda^{(1)})=\sigma^2_{in}+\sigma^2_{out}=2\sigma^2.
\label{eq:varsigma2}
\end{equation}

To solve for $\matr{u}^{(1)}$, we write it as a vector decomposition and solve for the individual components,
\begin{align}
    \matr{u}_1^{(1)} = \matr{u}_{\shortparallel}^{(1)} + \matr{u}_{\perp}^{(1)},
    \label{eq:u1decomp}
\end{align}
where $\matr{u}_{\shortparallel}^{(1)}$ is the component of $\matr{u}^{(1)}$ that is in the $\matr{u}_C, \matr{u}_H$ plane, and $\matr{u}_{\perp}^{(1)}$ is the component orthogonal to that plane. We find $\matr{u}_{\shortparallel}^{(1)}$ and $\matr{u}_{\perp}^{(1)}$ by multiplying both sides of the $\mathcal{O}(\sigma/\sqrt{N})$ equation, Eq.~(\ref{eq:1st_order}) by $\matr{u}_H^T$, the transpose of the homogeneous eigenvector of $\langle \matr{A} \rangle$, which gives
\begin{equation}
    \matr{u}_H^T(\langle \matr{A} \rangle \matr{u}^{(1)} + \matr{X} \matr{u}_C) = \matr{u}_H^T (\nu N \matr{u}^{(1)} + \lambda^{(1)} \matr{u}_C),
\end{equation}
which after employing the eigenvector properties becomes
\begin{equation}
    \mu N \matr{u}_H^T \matr{u}^{(1)} + \matr{u}_H^T  \matr{X} \matr{u}_C = \nu N \matr{u}_H^T \matr{u}^{(1)}.
\end{equation}
Rearranging and noting that $\matr{u}_H^T \matr{u}^{(1)} = \matr{u}_H^T \matr{u}_{\shortparallel}^{(1)}$, we find
\begin{equation}
    \matr{u}_H^T \matr{u}^{(1)} = \matr{u}_H^T \matr{u}_{\shortparallel}^{(1)} =  \frac{\matr{u}_H^T \matr{X} \matr{u}_C}{\nu N - \mu N},
\end{equation}
which suggests the following solution for $\matr{u}_{\shortparallel}^{(1)}$,
\begin{equation}
    \matr{u}_{\shortparallel}^{(1)} = \frac{[\matr{X} \matr{u}_C]_{\shortparallel}}{\nu N - \mu N}.
\end{equation}
Writing $[\matr{X}\matr{u}_C]_{\shortparallel}$ as a decomposed projection onto $\matr{u}_C$ and $\matr{u}_H$,
\begin{equation}
    [\matr{X}\matr{u}_C]_{\shortparallel} = (\matr{u}_C^T\matr{X}\matr{u}_C)\matr{u}_C + (\matr{u}_H^T\matr{X}\matr{u}_C)\matr{u}_H
\end{equation}
allows us to write $\matr{u}_{\shortparallel}^{(1)}$ as
\begin{equation}
    \matr{u}_{\shortparallel}^{(1)} = \frac{(\matr{u}_C^T\matr{X}\matr{u}_C)}{\nu N - \mu N}\matr{u}_C + \frac{(\matr{u}_H^T\matr{X}\matr{u}_C)}{\nu N - \mu N}\matr{u}_H.
    \label{eq:u1ParallelSol}
\end{equation}
We now seek to solve for $\matr{u}_{\perp}^{(1)}$. Using the decomposition (\ref{eq:u1decomp}) in the first order equation, (\ref{eq:1st_order}), gives
\begin{equation}
    \langle \matr{A} \rangle[\matr{u}_{\shortparallel}^{(1)} + \matr{u}_{\perp}^{(1)}] + \matr{X} \matr{u}_C = \nu N [\matr{u}_{\shortparallel}^{(1)} + \matr{u}_{\perp}^{(1)}] + \lambda^{(1)}\matr{u}_C.
\end{equation}
Retaining only the terms that have components orthogonal to the $\matr{u}_C, \matr{u}_H$ plane and rearranging yields
\begin{equation}
    \matr{u}_{\perp}^{(1)} = \frac{[\matr{X} \matr{u}_C]_{\perp}}{\nu N}.
\end{equation}
Using the substitution $[\matr{X} \matr{u}_C]_{\perp} = \matr{X} \matr{u}_C - [\matr{X} \matr{u}_C]_{\shortparallel}$ then gives the solution for the orthogonal component
\begin{equation}
    \matr{u}_{\perp}^{(1)} = \frac{\matr{X} \matr{u}_C - [(\matr{u}_C^T \matr{X}\matr{u}_C)\matr{u}_C + (\matr{u}_H^T \matr{X}\matr{u}_C)\matr{u}_H]}{\nu N}.
        \label{eq:u1PerpSol}
\end{equation}
Combining Eqs.~(\ref{eq:u1ParallelSol}) and (\ref{eq:u1PerpSol}) gives the solution for the first order perturbation to the eigenvector $\matr{u}^{(1)}$,
\begin{align}
\label{eq:v^1}
    \matr{u}^{(1)} = \frac{\matr{X}\matr{u}_C}{\nu N} + \frac{\mu (\matr{u}_C^T\matr{X}\matr{u}_C)}{\nu N(\nu - \mu)}\matr{u}_C + \frac{\mu (\matr{u}_H^T \matr{X}\matr{u}_C)}{\nu N(\nu - \mu)}\matr{u}_H.
\end{align}
\subsection{Second Order Treatment} \label{sec:secondorder}

Having found the first order eigenvalue and eigenvector perturbations, $\lambda^{(1)}$ and $\matr{u}^{(1)}$, we can now solve for the second order correction $\lambda^{(2)}$.  We multiply both sides of Eq.~(\ref{eq:2nd_order}) by $\matr{u}_C^T$ and then solve to get
\begin{align}
    \lambda^{(2)} &= \matr{u}_C^T \matr{X} \matr{u}^{(1)} - \lambda^{(1)} \matr{u}_C^T \matr{u}^{(1)}\\
    \label{eq:lam^2_allterms}
     &= \frac{\matr{u}_C^T\matr{X}^2\matr{u}_C}{\nu N} - \frac{(\matr{u}_C^T\matr{X}\matr{u}_C)^2}{\nu N} + \frac{\mu (\matr{u}_H^T\matr{X}\matr{u}_C)^2}{\nu N(\nu-\mu)},
\end{align}
where Eqs.~(\ref{eq:xux}) and (\ref{eq:v^1}) have been used to obtain the second line.

We seek the expected value $\langle \lambda^{(2)} \rangle$ and consider the righthand terms of Eq.~(\ref{eq:lam^2_allterms}) in succession. Expanding the expected value of the first term yields
\begin{equation}
\begin{split}
  \frac{ \langle  \matr{u}_C^T \matr{X}^2 \matr{u}_C \rangle}{\nu N} =
   \frac{1}{\nu N} \frac{1}{N} \Biggl\{ \sum_{i,j = 1}^{\frac{N}{2}} \langle (\matr{X}^2)_{ij} \rangle
    + \sum_{i,j = \frac{N}{2}+1}^{N} \langle (\matr{X}^2)_{ij} \rangle \\
    - 2 \sum_{i = 1}^{\frac{N}{2}}\sum_{j=\frac{N}{2}+1}^{N} \langle (\matr{X}^2)_{ij} \rangle \Biggl\},
    \label{eq:uX2u}
\end{split}
\end{equation}
where $(\matr{X}^2)_{ij} = \sum_{k=1}^N X_{ik} X_{kj}$. As the elements of $\matr{X}$ are independent, the cross-element terms in this sum have vanishing expectation: $\langle X_{ik} X_{kj} \rangle = 0$ for $i \neq j$. When $i = j$, the value of $\langle X_{ik}^2 \rangle$ is either the ingroup or outgroup variance: $\langle X_{ik}^2 \rangle = \sigma_{in}^2$ if $i,k \leq N/2$ or $i,k > N/2$; $\langle X_{ik}^2 \rangle = \sigma_{out}^2$ otherwise. Accordingly, the expectations for the elements of $\matr{X}^2$ are given by
\begin{align}
\langle (\matr{X}^2)_{ij} \rangle =
    \left\{
      \begin{array}{@{}lll@{}}
        \frac{N}{2}\sigma_{in}^2 + \frac{N}{2}\sigma_{out}^2 & = N \sigma^2, & i = j \\
        & & \\
        0, & & i \neq j .
      \end{array}\right.
\end{align}
The above equation reduces the double sums in the first two terms in Eq.~(\ref{eq:uX2u}) to single sums over $\langle (\matr{X}^2)_{ii} \rangle=N \sigma^2$, which can then be combined. The last term, which contains only off-diagonal elements of $\langle (\matr{X}^2) \rangle$, vanishes. The contribution of the first term in Eq.~(\ref{eq:lam^2_allterms}) to $\langle \lambda^{(2)} \rangle $ is therefore
\begin{align}
    \frac{\langle \matr{u}_C^T  \matr{X}^2 \matr{u}_C \rangle}{\nu N} &=\frac{1}{\nu N} \frac{1}{N} \sum_{i=1}^N N \sigma^2  \\
     &= \frac{\sigma^2}{\nu}. \label{eq:sigma2nu}
\end{align}

We now turn to the second and third terms on the righthand of Eq.~(\ref{eq:lam^2_allterms}). The numerator of the second term, $\ (\matr{u}_C^T\matr{X}\matr{u}_C)^2 $, involves the square of $\lambda^{(1)}$ by Eq.~(\ref{eq:xux}). Hence, its expected value is equivalent to the variance of $\lambda^{(1)}$ (which has zero mean), and so goes as $\sigma^2$ as shown above. Consequently, the expected value of the second term goes as $\sigma^2/N$. The same argument holds for the third term as its numerator depends on $\matr{u}_H^T\matr{X}\matr{u}_C$, which is likewise tantamount to the sum over the random ingroup and outgroup variables. Therefore, in the large-$N$ regime of concern here, the second and third terms, which have a $1/N$ dependence, can be neglected in comparison with the first term, which is independent of $N$. Accordingly, Eq.~(\ref{eq:sigma2nu}) gives the second order eigenvalue perturbation
\begin{align}
    \langle \lambda^{(2)} \rangle =\frac{\sigma^2}{\nu}.
    \label{eq:lam2}
\end{align}

Having found that the first order perturbation vanishes on average and given the second order perturbation above, we arrive at the approximate solution for the expected value of the contrast eigenvalue of Eq.~(\ref{eq:eigdef}),
\begin{align}
\label{eq:lamFpert}
    \langle \lambda_C \rangle &= \nu N + \frac{\sigma^2}{\nu}.
\end{align}
Since the ratio of the second-order correction to the unperturbed eigenvalue goes as $\sigma^2/N$, the expansion parameter, given by its square root, is therefore $\mathcal{O}(\sigma/\sqrt{N})$ as stated at the beginning of this calculation.

A similar calculation, this time expanding about the homogeneous eigenvector $\matr{u}_H$, gives us the solution for $\lambda_H$, which simply involves swapping out $\nu$ for $\mu$ in the preceding equation,
\begin{align}
\label{eq:lamHpert}
    \langle \lambda_H \rangle &= \mu N + \frac{\sigma^2}{\mu}.
\end{align}

\begin{figure}[!htbp]
\centering
\begin{tikzpicture}[scale=1]
\node[inner sep=0pt](russell) at (0,0)
 {\includegraphics[width=0.95\linewidth]{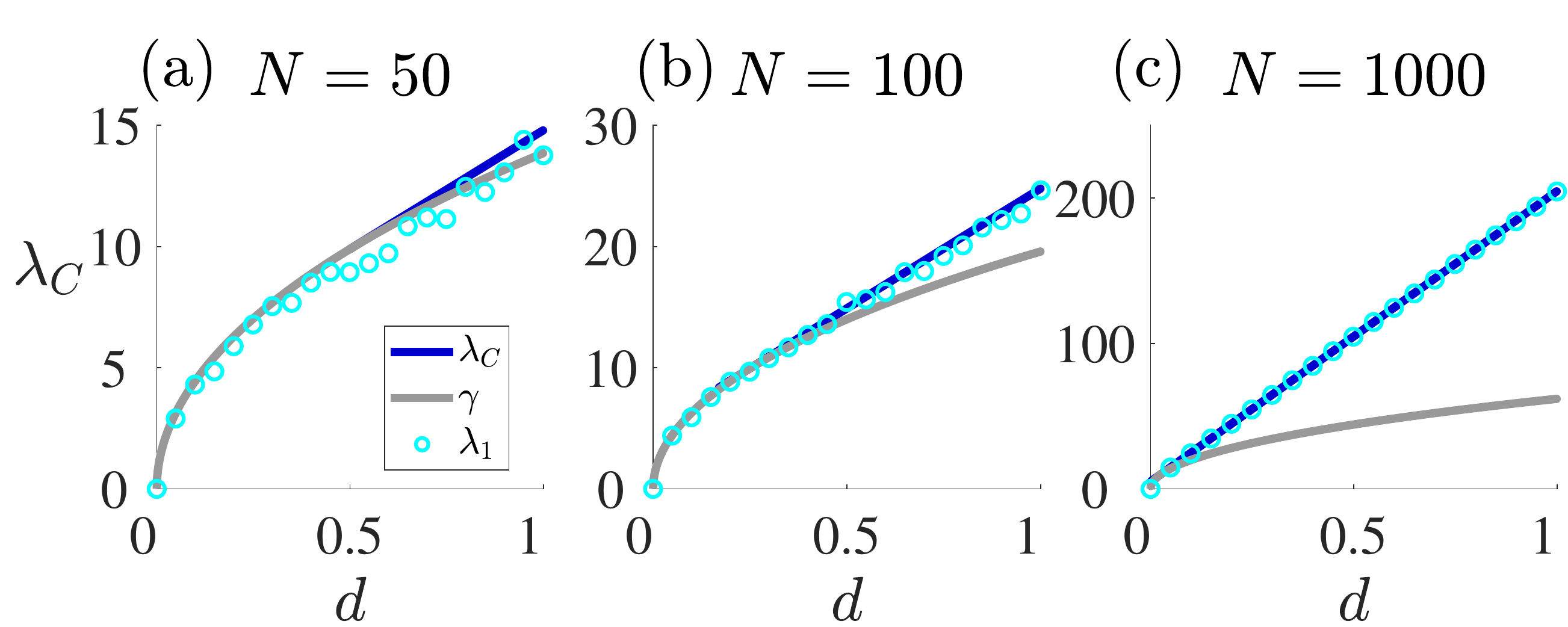}};
\end{tikzpicture}
\caption{Theoretical contrast ($\lambda_C$, Eq.~(\ref{eq:lamFpert})) and band edge ($\gamma$, Eq.~(\ref{eq:gamma2})) eigenvalues, along with the simulated leading eigenvalue ($\lambda_1$), as a function of network density. (a) $N=50$ (b) $N=100$ (c) $N=1000$. $\lambda_C$ is plotted for network density values where it meets or exceeds $\gamma$. One leading eigenvalue instance is computed for each $d$ value. $d = d_{in} = d_{out}$, $p_{in}^+ = 0.6$, and $p_{out}^+ = 0.4$. }
\label{fig:elaboration}
\end{figure}

The analytical expressions for the signal eigenvalues, (\ref{eq:lamFpert}) and (\ref{eq:lamHpert}), are plotted in Fig.~\ref{fig:zH_zF_spectra}(a) for values outside the main spectral band. They are observed to be in good agreement with the outlying eigenvalues of the numerical spectrum. This is the case even though, rather than an average over many generated networks, each horizontal slice represents just one instance, a reflection of the self-averaging behavior of large random networks. These expressions also work well in the non-sparse limit as shown in Fig.~\ref{fig:elaboration} for the case where the contrast eigenvalue becomes the leading eigenvalue beyond the assortative transition. The predicted $\lambda_C$  is observed to separate from the spectral edge past a critical density, which decreases with network size, and shows good agreement with the first eigenvalue of the simulated network, particularly for the two larger networks.

The symmetric forms of the expressions for $\lambda_C$ and $\lambda_H$ reflect the fact that an orthogonal transformation $\matr{K}$ exists that transforms the contrast and homogeneous eigenvectors of $\langle \matr{A} \rangle$ into each other, that is, $\matr{u}_{H}^\prime = \matr{K}\matr{u}_{C}$ and $\matr{u}_{C}^\prime = \matr{K}\matr{u}_{H}$ where the primes denote the transformed system. Specifically, $\matr{K}$ is the diagonal matrix $\mathrm{diag}(1,\ldots,1,-1,\ldots,-1)$ where the negative values start at index $N/2+1$. It is its own inverse, $\matr{K}^{-1}=\matr{K}$. The transformation of the expected adjacency matrix, $\langle \matr{A}^\prime \rangle = \matr{K}\langle\matr{A}\rangle\matr{K}$, flips the sign of the off-diagonal blocks so that $\langle A_{out}^\prime \rangle = -\langle A_{out} \rangle$. Therefore, by Eqs.~(\ref{eq:mu}) and (\ref{eq:nu}), $\mu^\prime = \nu$ and $\nu^\prime = \mu$, which swaps the perturbed signal eigenvalues, $\lambda_{H}^\prime = \lambda_C$ and $\lambda_{C}^\prime = \lambda_H$.

An alternative calculation of the signal eigenvalues based on random matrix theory and complex analysis is presented in the Appendix~\ref{appendix2}. We find the same formulas for the signal eigenvalues (see Eq.~(\ref{eq:z_F_complex})) as have been derived here.

\section{\label{sec:trans_bound} Transition Boundaries}
In this section, we derive theoretical predictions for the boundaries of the detectability and sociality transitions. As discussed in Sec.~\ref{sec:detect}, these transitions occur when the signal eigenvalues merge with the main spectral band.  From Eq.~(\ref{eq:gamma2}), the edges of the main band of $\matr{X}$ are given by $\pm 2\sigma\sqrt{N}$, a formula that is a straightforward adaptation of the band edge of Wigner's semicircle distribution. We consider the community detectability transitions first, that is, those involving the contrast eigenvector $\matr{u}_C$, whose eigenvalue is given by Eq.~(\ref{eq:lamFpert}). The detectability transition will therefore occur when
\begin{equation}
    \nu N + \frac{\sigma^2}{\nu} = 2\sigma \sqrt{N}.
\end{equation}

Solving for the critical value $\nu_*$ yields
\begin{equation}
\nu_* = \frac{\sigma}{\sqrt{N}}.
\label{eq:nucritical}
\end{equation}
We note that $\sigma$ depends upon $\nu$ as given by Eq.~(\ref{eq:sigma2munu}). The community structure is detectable when $|\nu| > \nu_*$. In particular, the assortative transition occurs for $\nu=\nu_*$ and the detectability transition for disassortative structure occurs for $\nu=-\nu_*$.

We observe that the transition condition (\ref{eq:nucritical}) also results by setting the noise power equal to the signal power. Defining the noise power as the projection of $\matr{X}^2$ onto $\matr{u}_C$, its average, $\langle \matr{u}_{C}^{T} \matr{X}^2 \matr{u}_C \rangle$, is found using Eq.~(\ref{eq:sigma2nu}) to be $N\sigma^2$. One could also arrive at this value by considering how much of the total noise variance, $N^2\sigma^2$, is carried on average by each of $N$ randomly chosen orthogonal basis vectors. Equating the signal power to the average noise power, $\nu^2 N^2=N\sigma^2$, yields (\ref{eq:nucritical}).

The sociality transitions associated with the homogeneous signal occur when the homogeneous eigenvalue given by Eq.~(\ref{eq:lamHpert}) equals the band edge eigenvalue. This yields a critical value $\mu_*$,
\begin{equation}
\mu_* = \frac{\sigma}{\sqrt{N}}.
\label{eq:mucritical}
\end{equation}
The prosocial transition occurs for $\mu=\mu_*$ and the antisocial transition occurs for $\mu=-\mu_*$.

We now unpack the transition conditions derived above to express them in alternative ways in parameter space that will further intuitive understanding of the transition behavior and allow for connection with simulation results.

First, we substitute Eq.~(\ref{eq:sigma2munu}) for $\sigma^2$ in the detectability condition (\ref{eq:nucritical}) to yield
\begin{equation}
N\nu^2 = \frac{1}{2}(d_{in}+d_{out}) - \mu^2 -\nu^2.
\label{eq:nusq}
\end{equation}
We point out that $(d_{in}+d_{out})/2$ is simply the overall tie density in the network. For a sparse network, $d_{in}, d_{out} \ll 1$, we can neglect the $\mu^2$ and $\nu^2$ terms on the right-hand side, so that the critical value $\nu_*$ is given by
\begin{equation}
\label{eq:fact_sparse}
\nu_* = \pm \sqrt{\frac{1}{2N}(d_{in} + d_{out})}.
\end{equation}
The positive sign corresponds to the assortative transition and the negative sign corresponds to the disassortative transition. As $\nu$ can be regarded as a natural parameter for the community structure, this structure (assortative or disassortative) becomes easier to detect as the network becomes more sparse since $|\nu_*|$ shifts to smaller values (but care should be taken to distinguish the behavior of $\nu$ from that of the affinities and animosities, which can behave oppositely with density as in Eq.~(\ref{eq:comm_det_curve}) below). Weaker structure is also more detectable as the size of the network grows, as was already apparent from Eq.~(\ref{eq:nucritical}).

We now substitute into Eq.~(\ref{eq:nusq}) the definitions (\ref{eq:mu}) and (\ref{eq:nu}) for $\mu$ and $\nu$ and rearrange to obtain
\begin{align}
\begin{split}
0&=\langle A_{out} \rangle^2 - \frac{2N}{N+2} \langle A_{out} \rangle \langle A_{in} \rangle\\
&+ \langle A_{in} \rangle^2 -\frac{2}{N+2}(d_{in}+d_{out}),
\end{split}
\end{align}
which can be solved for the critical value of $\langle A_{out} \rangle$ (omitting the asterisk),
\begin{align}
\begin{split}
\langle A_{out} \rangle &= \frac{N}{N+2} \langle A_{in} \rangle \\  &\pm \sqrt{\frac{2}{N+2}(d_{in}+d_{out}) - \frac{4(N+1)}{(N+2)^2}\langle A_{in} \rangle^2}.
\end{split}
\label{eq:Aotrans}
\end{align}

To write the detectability transitions completely in terms of the block model probabilities, we substitute  Eqs.~(\ref{eq:avAi}) and (\ref{eq:avAo}) for $\langle A_{in} \rangle$ and $\langle A_{out} \rangle$ into (\ref{eq:Aotrans}). Solving for the outgroup animosity and taking the large $N$ limit yields
\begin{align}
\label{eq:fact_trans_curve}
\begin{split}
p_{out}^- &= \frac{1}{2} - \frac{d_{in}}{d_{out}} \biggl(p_{in}^+ - \frac{1}{2} \biggl)\\
&\pm \frac{1}{d_{out}} \sqrt{\frac{d_{in} + d_{out} - 8d_{in}^2(p_{in}^+-\frac{1}{2})^2}{2N}},
\end{split}
\end{align}
where the positive sign corresponds to the assortative transition. Neglecting the second term inside the square root yields the sparse limit, equivalent to Eq.~(\ref{eq:fact_sparse}).

For the sociality transitions, the sparse limit results in the condition
\begin{equation}
\label{eq:homog_sparse}
\mu_* = \pm \sqrt{\frac{1}{2N}(d_{in} + d_{out})}.
\end{equation}
The positive and negative signs correspond to the prosocial and antisocial transitions respectively. The critical value of the outgroup animosity is given by
\begin{align}
\label{eq:homog_trans_curve}
\begin{split}
p_{out}^- &= \frac{1}{2} + \frac{d_{in}}{d_{out}} \biggl(p_{in}^+ - \frac{1}{2} \biggl)\\
&\pm \frac{1}{d_{out}} \sqrt{\frac{d_{in} + d_{out} - 8d_{in}^2(p_{in}^+-\frac{1}{2})^2}{2N}},
\end{split}
\end{align}
where now the negative sign is used for the prosocial transition.
\\
\\

%
%%%%%%%%%%%%%%%%%%%%%%%%%%%%%%
%%%%% new section %%%%%%%%%%%%
%%%%%%%%%%%%%%%%%%%%%%%%%%%%%%
\section{\label{sec:StrucBalDyn}Structural Balance Dynamics}
In its simplest incarnation, structural balance theory considers the stability of triads.  Triads with all positive ties (``the friend of my friend is my friend'') or two negative ties (``the enemy of my enemy is my friend") are considered balanced and so stable. In contrast, a triad with an odd number of negative ties will be unbalanced. For fully connected networks, assuming that all triads must be balanced over time implies that the system achieves either a state of global harmony in which all nodes are positively connected or two hostile camps with positive connections within each camp and negative connections between them \cite{CarHar1956}. Empirical signed networks in social systems such as international relations, student relationships, and online social networks have been found to be approximately balanced \cite{kirkley_balance_2019,traag_partitioning_2018,facchetti_computing_2011}, exhibiting a tendency toward partition into two factions.

\par
Although the concept of balance can be extended to arbitrary-length cycles, the triadic notion has motivated the construction of dynamical systems models that evolve the relationship between a pair of nodes as a function of their relationships with their network neighbors \cite{KulGawGro2005,marvel_continuous-time_2011,traag_dynamical_2013}. If both members of a dyad have a positive relationship with a third node, that will act towards making the focal dyad's relationship more positive. In contrast, having oppositely signed relationships with the third node will contribute a force pulling the dyad toward a more conflictual relationship. Prompted by this notion that unbalanced triads generate tension in social networks and building upon Ref.~\cite{KulGawGro2005}, Marvel et al. \cite{marvel_continuous-time_2011} demonstrated that the following model of structural balance dynamics almost always achieves a balanced state starting from random initial conditions:
\begin{equation} \label{eq:cont time model}
    \frac{d\matr{Y}}{dt} = \matr{Y}^2,
\end{equation}
where $t$ is time and $\matr{Y}_{ij}$ is the connectivity value from node $i$ to node $j$. The connectivity matrix $\matr{Y}$ refers to the dynamically evolving network containing signed and continuous tie values. For the purposes of analyzing community structure, we will convert $\matr{Y}$ to the adjacency matrix $\matr{A}$ with discrete values $\pm 1$ and 0 by taking the sign of the connectivity values. In support of its empirical relevance, Ref.~\cite{marvel_continuous-time_2011} found that, when implemented upon the initial network of several real world systems, this model well predicts the observed final network.
\par
Equation~(\ref{eq:cont time model}) is the matrix form of a Riccati equation and has the following closed form solution for $t<t^*$, where $t^*$ is the time at which solutions go to positive and negative infinity \cite{marvel_continuous-time_2011}:
\begin{equation} \label{eq:soln riccati}
    \matr{Y}(t) = \matr{Y}(0)[I - \matr{Y}(0)t]^{-1}.
\end{equation}
As $t \to t^*$, the leading eigenvector of the initial network, $\matr{Y}(0)$, dominates the connectivity structure. As a result, the final adjacency matrix corresponding to $\matr{Y}$ as $t \to t^*$ converges to the outer product
\begin{equation} \label{eq:final matrix soln}
    \matr{A} = \matr{u}_1 \matr{u}_1^T,
\end{equation}
where $\matr{u}_1$ consists of the signs of the leading eigenvector of $\matr{Y}(0)$. Thus, the rank 1 structure toward which the connectivity matrix converges implies that the network must partition into either two hostile factions or one harmonious community as consistent with the expectations of structural balance theory \cite{marvel_continuous-time_2011}. The final network consists of a single harmonious faction if the components of $\matr{u}_1$ are of uniform sign, but consists of two hostile factions if $\matr{u}_1$ contains both positive and negative values \cite{marvel_continuous-time_2011}.
%
%% tikz picture %%
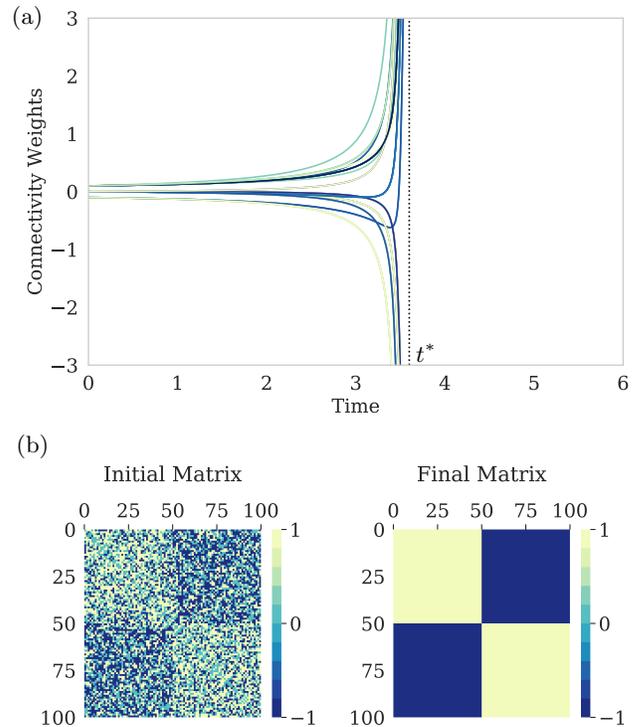
\begin{figure}[!htbp]
\begin{tikzpicture}[scale=1]
\node[inner sep=0pt](russell) at (0,0)
 {\includegraphics[width=0.95\linewidth]{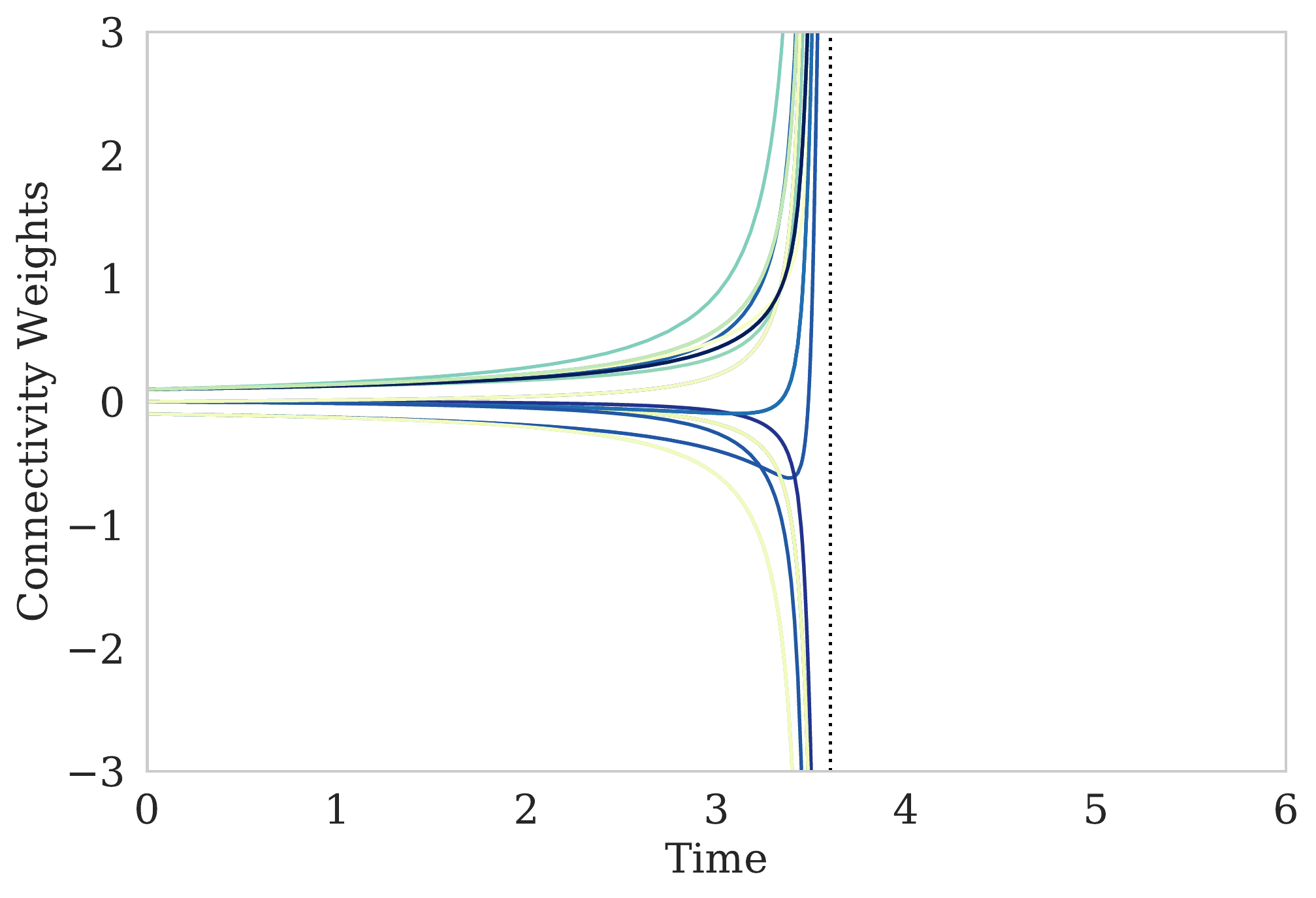}};
 \node at (-4,2.6) {\small (a)};
 \node at (1.3,-1.85) {$t^*$};
\end{tikzpicture}
\begin{tikzpicture}[scale=1]
\node[inner sep=0pt](russell) at (0,0)
 {\includegraphics[width=0.92\linewidth]{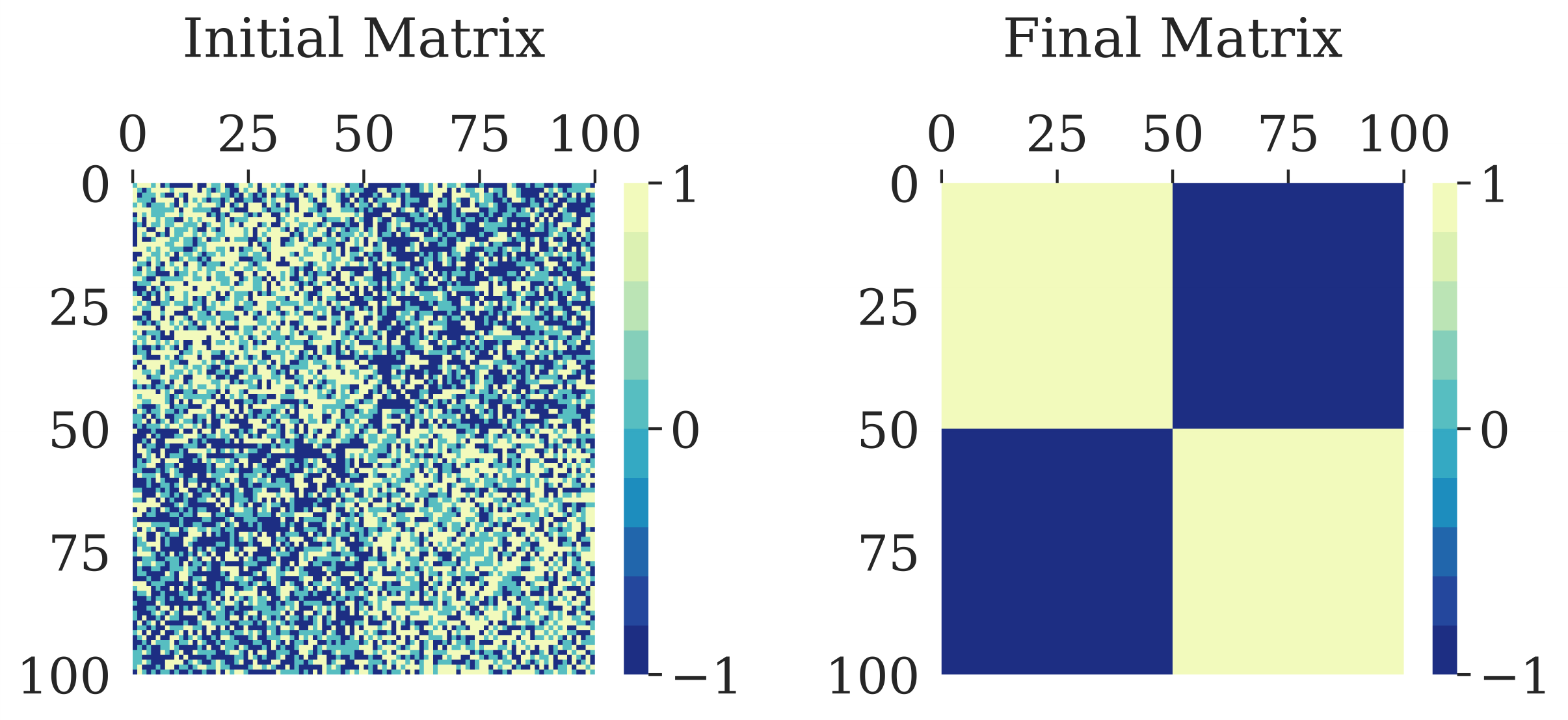}};
 \node at (-4,2) {\small (b)};
 \end{tikzpicture}
\caption{Structural balance evolution of a network with community structure. (a) Connectivity network weights $\matr{Y}_{ij}(t)$ over time evolved by Eq.~(\ref{eq:cont time model}) from an initial network generated by a stochastic block model (with values rescaled by $1/N$. (b) Initial and final adjacency matrices. Initial matrix parameters $N=100$, $d_{in} = 0.7$, $d_{out} = 0.7$, $p_{in}^+ = 0.65$, and $p_{out}^+ = 0.35$.}
\label{fig:struc_balance}
\end{figure}
\par
To illustrate its application to networks with initial community structure, Fig.~\ref{fig:struc_balance}(a) shows the evolution of network connectivity values over time from a $\matr{Y}(0)$ that is proportional to the adjacency matrix generated by the stochastic block model (Fig.~\ref{fig:struc_balance}(b), left). The final adjacency matrix shows the split into two factions with positive ties within each faction and negative ties between faction (Fig.~\ref{fig:struc_balance}(b), right). It is given by the outer product (\ref{eq:final matrix soln}) with $\matr{u}_1 = \matr{u}_C$ so that the factions correspond with the identity blocks $A$ and $B$.

%% new section %%%
\section{\label{sec:regimes}Structural Balance Behavioral Regimes}
As the final structure of these dynamical networks is dominated by the initial network's leading eigenvector, we can determine the extent to which networks in our parameter space will become assortative or homogeneous using our transition formulas derived above. First we treat a special case before exploring more general parameters.
\subsection{\label{sec:sign_sym}Ingroup affinity equals outgroup animosity}
We  consider the simple case in which the ingroup affinity is set equal to the outgroup animosity, $p_{in}^+ = p_{out}^-$. We also make the simplification $d_{in} = d_{out} =d$. For this case, $\langle A_{in} \rangle=-\langle A_{out} \rangle$ so that $\mu = 0$, i.e., there is no homogeneous signal, and $\nu = -\langle A_{out} \rangle$. Using $\langle A_{out} \rangle = d(1-2p_{out}^-)$ in Eq.~(\ref{eq:fact_sparse}), we solve for the critical outgroup animosity
\begin{equation}
    \label{eq:comm_det_curve}
    p_{out}^- = \frac{1}{2}\left(1 + \sqrt{\frac{1}{dN}}\right).
\end{equation}
Note that we only use the positive sign from (\ref{eq:fact_sparse}) since it is the assortative transition that involves the leading eigenvector.

%
%% tikz %%
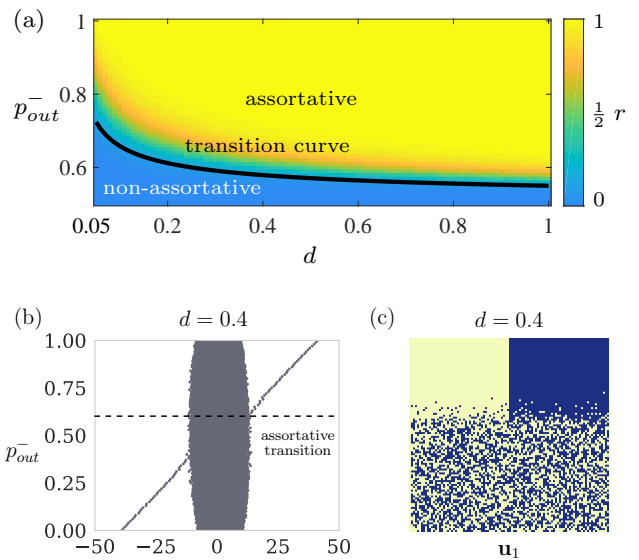
\begin{figure}[!htbp]
    \centering
 \subfloat{
\centering
\resizebox{\linewidth}{!}{
    \begin{tikzpicture}[scale=1]
        \node[inner sep=0pt](russell) at (0,0)
        {\includegraphics[width=0.75\linewidth]{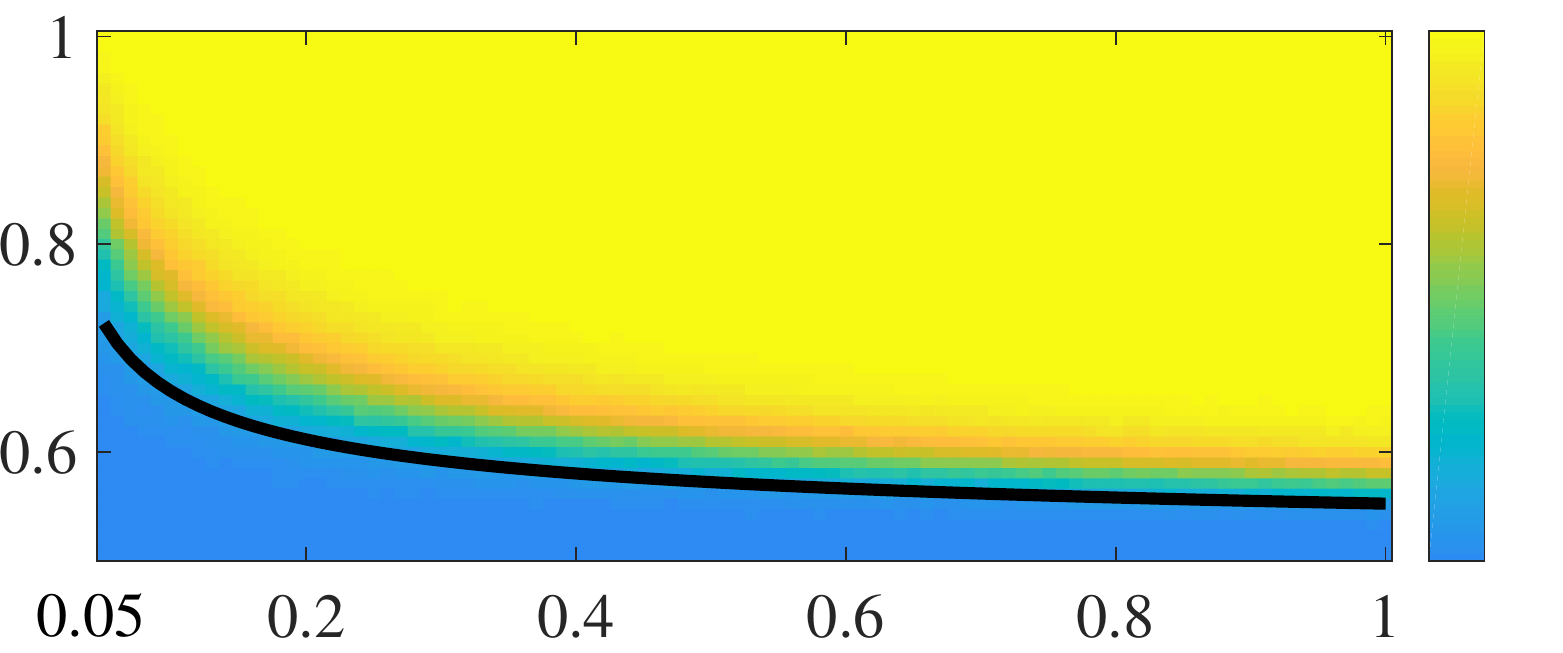}};
        \node at (-3.6,1.2) {\footnotesize (a)};
        \node at (-3.5,0.2) {\footnotesize $p_{out}^-$};
        \node at (-0.3,-1.55) {\footnotesize $d$};
        \node at (-0.4,0.3) {\scriptsize assortative};
        \node at (-0.8,-0.25) {\scriptsize transition curve};
        \node[white] at (-1.8,-0.75) {\scriptsize non-assortative};

        \node at (3.35,0.1) {\footnotesize $r$};
        \node at (3.1,1.2) {\scriptsize $1$};
        \node at (3.1,0.1) {\scriptsize $\frac{1}{2}$};
        \node at (3.1,-0.9) {\scriptsize $0$};
    \end{tikzpicture}
    }}
\vfill
\subfloat{
\centering
\resizebox{\linewidth}{!}{
\begin{tikzpicture}[scale=1]
        \node[inner sep=0pt](russell) at (0,0)
        {\includegraphics[width=0.7\linewidth]{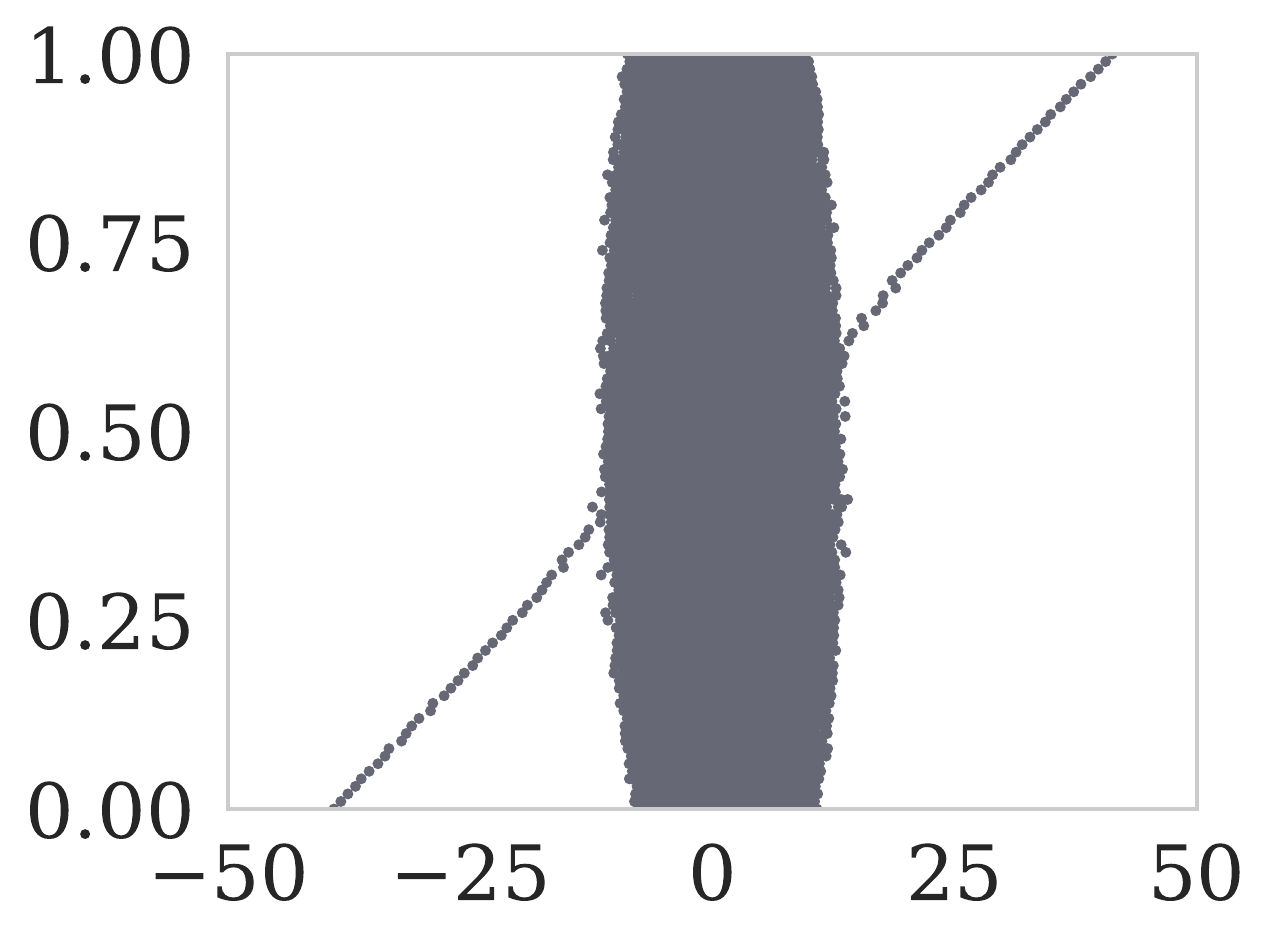}};
        \node at (-3.2,2.4) {\large (b)};
        \node at (0.3,2.4) {\large $d=0.4$};
        \node at (-3.3,-0.1) {\large $p_{out}^-$};
        \node at (1.9,0.2) {\footnotesize assortative};
 \node at (1.9,-0.1) {\footnotesize transition};
 \draw[dashed,thick](-1.95,0.55)--(2.68,0.55);
    \end{tikzpicture}
    \begin{tikzpicture}[scale=1]
        \node[inner sep=12pt](russell) at (0,0)
        {\includegraphics[width=0.47\linewidth]{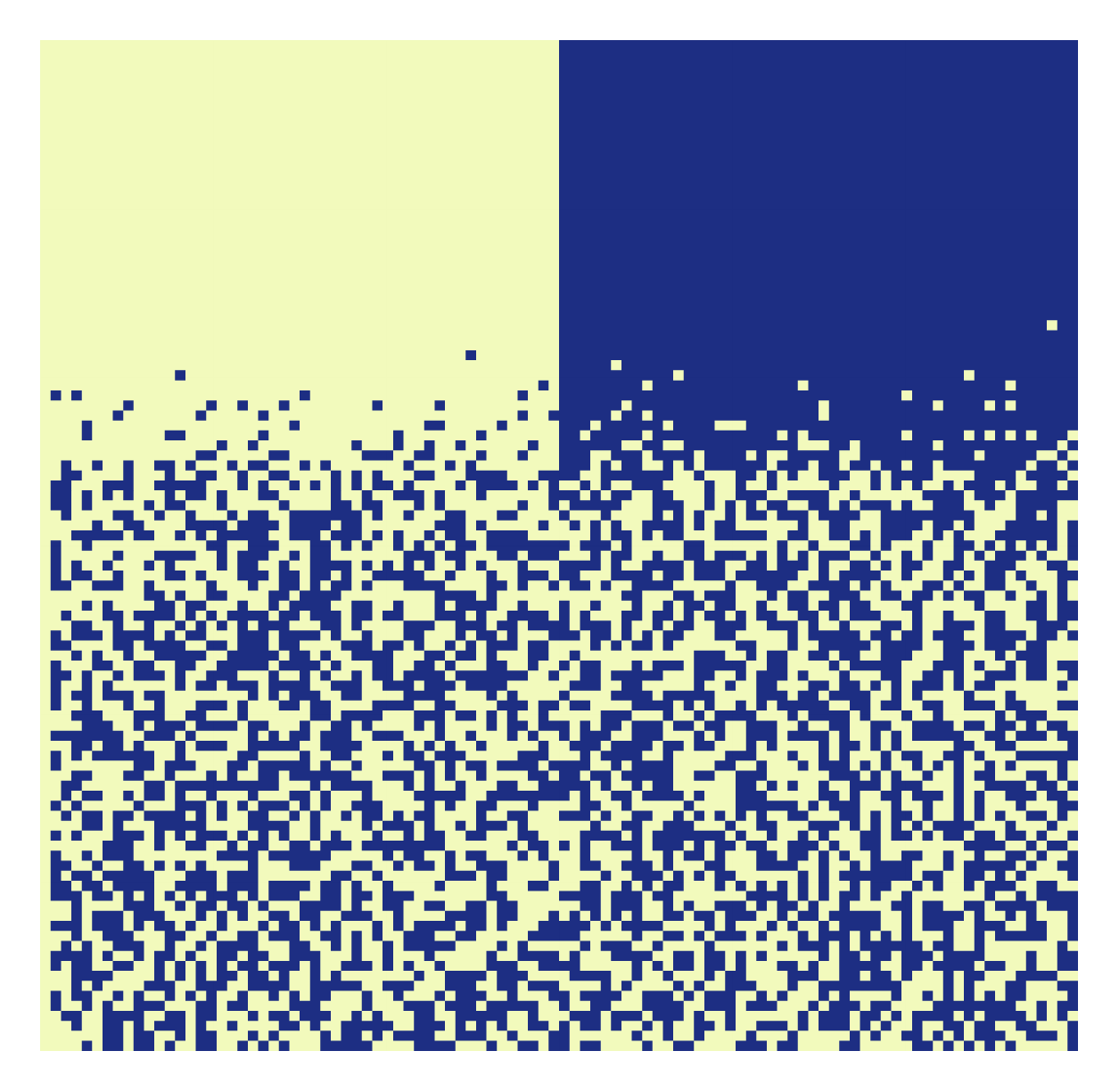}};
        \node at (-2.4,2.2) {\large (c)};
        \node at (0,2.2) {\large $d=0.4$};
        \node at (0,-2.2) {\large $\matr{u}_1$};
    \end{tikzpicture}
}}
    \caption{(a) Assortativity of final adjacency matrix as a function of $d$ and $p_{out}^-$ for network size $N=100$ averaged over 400 simulations. Solid curve is the theoretical transition boundary given by Eq.~(\ref{eq:comm_det_curve}). Note that the theoretical curve corresponds to the early part of the transition in which $r$ just begins to rise whereas the more visually distinctive yellow-cyan interface marks the middle of the transition. (b) Initial adjacency matrix  spectra for $d=0.4$ and increasing $p_{out}^-$. (c) Component signs for the leading eigenvector $\matr{u}_1$ of the initial adjacency matrix. The disassortative transition is not dynamically relevant so only the upper half of the $p_{out}^-$ scale is plotted in (a).}
    \label{fig:assort_heatmap}
\end{figure}
\par
Figure~\ref{fig:assort_heatmap} shows the alignment between the dynamical regimes evolved by the structural balance model and the community structure of the initial network. Panel (a) plots the assortativity $r$ defined by Eq.~(\ref{eq:assort}) of the final adjacency matrix as averaged over 400 initial networks generated by the stochastic block model at each point in the $d$ and $p_{out}^-$ parameter space. The yellow region represents the fully assortative outcome where the system evolves into two factions corresponding to the identities defined by the stochastic block model. In the blue region where $r \approx 0$, the two final factions are well mixed by identity. We see that the boundary between these two regions is in good accord with Eq.~(\ref{eq:comm_det_curve}) and, in particular, that denser networks become assortative at smaller outgroup animosity values so that the size of the non-assortative region decreases. Under the structural balance dynamics, the final network is determined by the leading eigenvector of $Y(0)$ by Eq.~(\ref{eq:final matrix soln}). Since community detectability is determined by the leading eigenvector as shown in Fig.~\ref{fig:assort_heatmap}(b) and (c) which depict the spectrum for a particular density value, the presence of spectrally detectable communities in the initial network induces assortativity in the final state. These plots also confirm that the leading eigenvector is never homogeneous and so one does not expect to observe the single faction outcome in this case (being extremely improbable).

%%%%%%%%%%%%%%%%%%%%%%%%%%%%%%%%%%%%%%%%%%%%%%%
\subsection{\label{sec:level1}General parameter conditions}
We now analyze the final states of networks generated using more general parameter conditions. We plot the behavior of the signal transition curves in the two-dimensional parameter space defined by $p_{in}^+$ and $p_{out}^-$ for fixed values of $d_{in}/d_{out}$.

As the homogeneous signal was irrelevant in the previous case, we needed only plot the assortativity $r$. However, the prosocial transition will occur in general and so we must measure the extent to which nodes can be found in one large group. We define the homogeneity $h$ as the fraction of all nodes that can be assigned to a single group by virtue of having a common sign in the leading eigenvector of the adjacency matrix. When all nodes have a common sign, they are positively connected to all other nodes so that $h = 1$ while when nodes are divided into two equal factions, the homogeneity assumes its minimum value, $h = 1/2$.

\par
\begin{figure}[!htbp]
\subfloat{
\centering
\resizebox{\linewidth}{!}{
    \begin{tikzpicture}[scale=1]
        \node[inner sep=0pt](russell) at (0,0)
        {\includegraphics[width=0.7\linewidth]{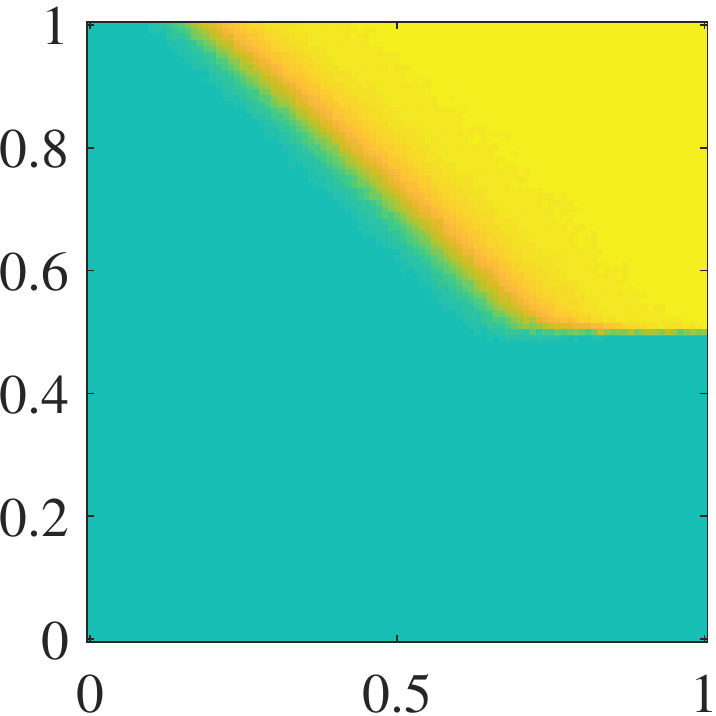}};
        \node at (-0.2,3.3)  {\Large (a) \     \ Assortativity};
        \node at (-3.5,0.1)  {\Large $p_{out}^-$};
        \node at (0.4,-3.4)  {\Large $p_{in}^+$};
        \node at (1.5,1.5)  {\Large $r=1$};
        \node at (0,-1.5)  {\Large $r=0$};
        \draw[line width=0.5mm,->] (1,-3)--(1.5,-4);
    \end{tikzpicture}
    \begin{tikzpicture}[scale=1]
        \node[inner sep=0pt](russell) at (0,0)
        {\includegraphics[width=0.7\linewidth]{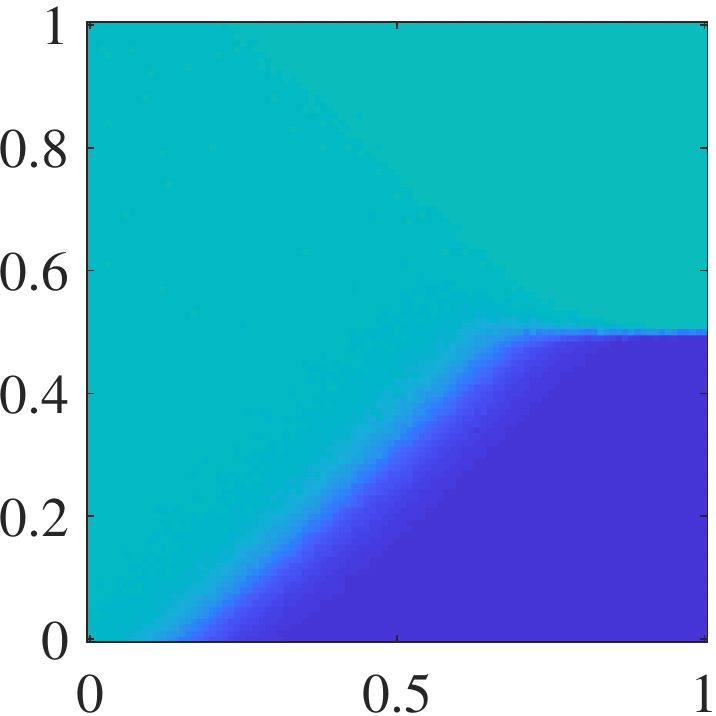}};
        \node at (-0.2,3.3)  {\Large (b) \     \ Homogeneity};
        \node at (0.4,-3.4)  {\Large $p_{in}^+$};
        \node[white] at (1.5,-1.5)  {\Large $h=1$};
        \node at (0,1.5)  {\Large $h=\frac{1}{2}$};
        \draw[line width=0.5mm,->] (-1,-3)--(-1.5,-4);
    \end{tikzpicture}
    }}
    \vfill
\subfloat{
\centering
\resizebox{\linewidth}{!}{
    \begin{tikzpicture}[scale=1]
        \node[inner sep=0pt](russell) at (-0.0,0)
        {\includegraphics[width=0.75\linewidth]{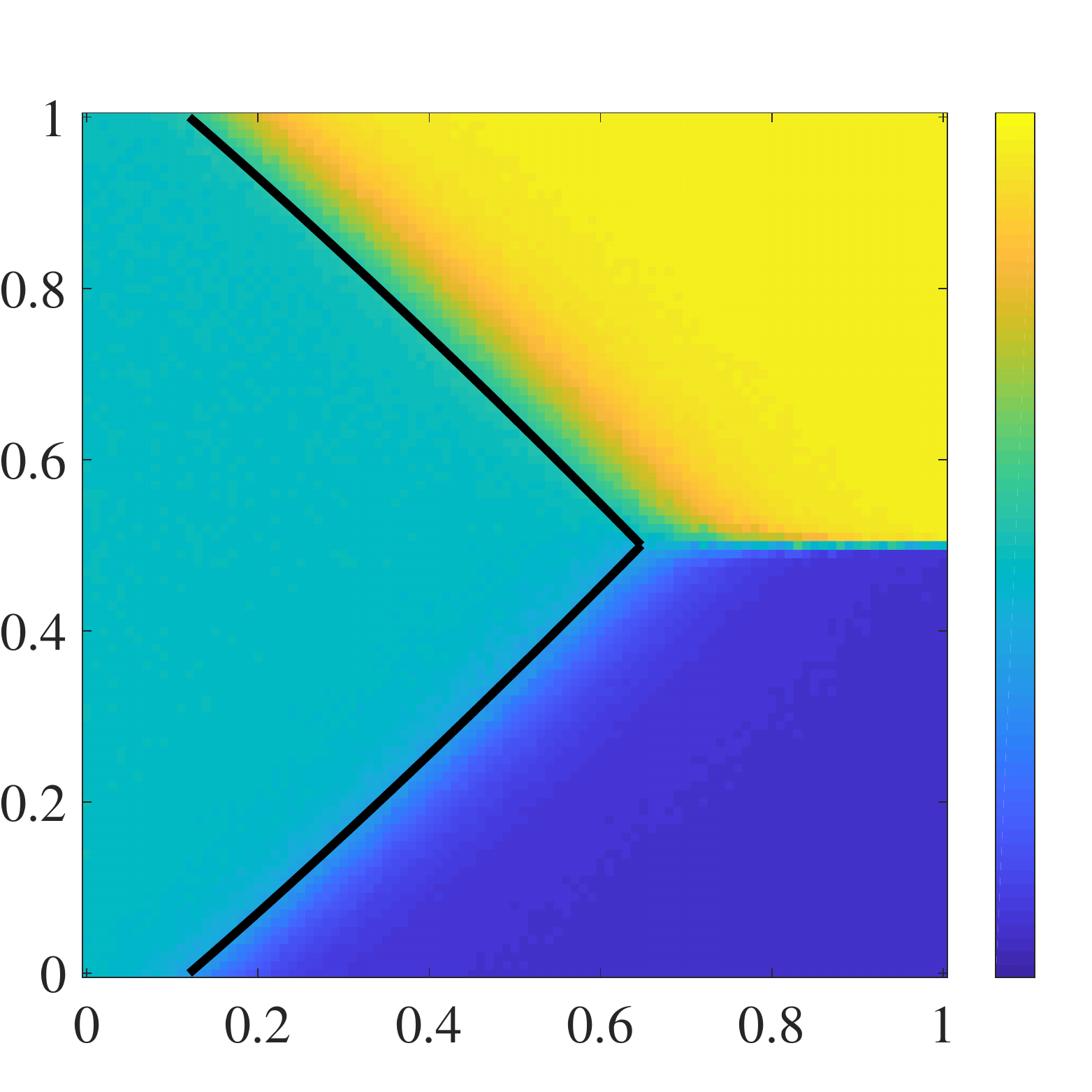}};
        \node at (-0.8,2.9) {(c) \ \ \ \ \ \ \ \ \ \ \ \ \  $d_{in}/d_{out} = 1$};
        \node at (1,2.1) {2 factions};
         \node at (1,1.7) {assortative};
        \node at (1,1.3) {$r = 1$};
        \node at (1,0.8) {$h = \frac{1}{2}$};
        \node[white] at (1,-0.6) {1 faction};
         \node[white] at (1,-1) {homogeneous};
        \node[white] at (1,-1.5) {$h = 1$};
        \node[white] at (1,-2) {$r = 0$};

        \node at (-1.4,0.8)  {2 factions};
        \node at (-1.4,0.4)  {non-assortative};
        \node at (-1.4,0)  {non-homogeneous};
        \node at (-1.4,-0.55)  {$h = \frac{1}{2}$};
        \node at (-1.4,-1.05) {$r = 0$};
        \node at (-3.65,0) {$p_{out}^-$};
        \node at (-0.1,-3.4) {$p_{in}^+$};

        \node at (3.1,2.4)  {$1$};
        \node at (3.1,0.3)  {$0$};

        \node at (3.1,-0.3)  {$\frac{1}{2}$};
        \node at (3.1,-2.4)  {$1$};

        \draw[line width=0.25mm,|->] (3.4,-0.1)--(3.4,-2.5);
        \draw[line width=0.25mm,|->] (3.4,0.1)--(3.4,2.5);

        \node at (3.6,1.2)  {$r$};
        \node at (3.6,-1.2)  {$h$};

        \draw [fill=white,white] (2.6,-0.1) rectangle (3,0.1);
        \draw[line width=0.2mm,-] (2.66,0.1)--(2.92,0.1);
        \draw[line width=0.2mm,-] (2.66,-0.1)--(2.92,-0.1);
        \end{tikzpicture}
        }
        }
        \caption{ (a) Assortativity and (b) homogeneity of final network states evolved by structural balance model (\ref{eq:cont time model}) as a function of ingroup affinity and outgroup animosity. (c) Assortativity and homogeneity are mapped using the measure $z = r - 2h +1$ to generate the joint heat map. For convenience, two separate color bar scales are shown instead of z. The upper black curve indicates the assortative transition boundary, Eq.~(\ref{eq:fact_trans_curve}), while the lower black curve indicates the homogeneous transition boundary, Eq.~(\ref{eq:homog_trans_curve}). Heatmap values generated by averaging over $20$ simulations for parameters $d_{in} = d_{out} = 0.45$ and $N=100$. }
       \label{fig:heatmap_gen1}
\end{figure}
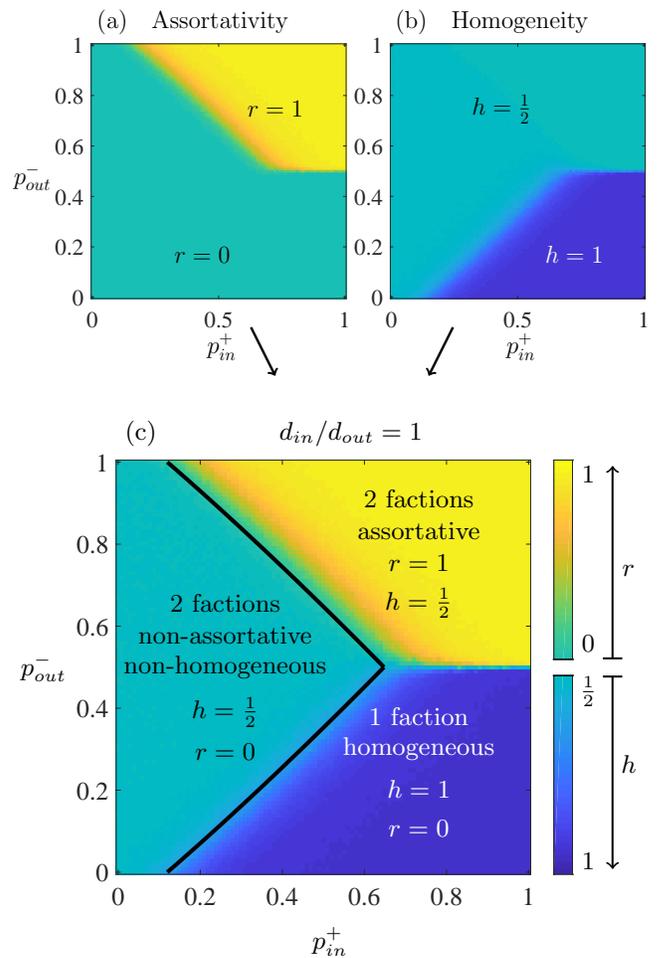

The top plots of Fig.~\ref{fig:heatmap_gen1} show the assortativity and homogeneity of the final adjacency matrix evolved by the structural balance dynamics in the parameter space defined by the ingroup affinity and outgroup animosity. They can then be linearly combined to effect their joint visualization as shown in the bottom plot. The assortative transition boundary predicted by Eq.~(\ref{eq:fact_trans_curve}) separates the assortative from non-assortative two-faction states while the prosocial theoretical boundary of Eq.~(\ref{eq:homog_trans_curve}) separates homogeneous single-faction states from the non-assortative two-faction states. These regimes relate to the initial network spectrum as follows: the blue region is where the homogeneous eigenvalue is both the largest eigenvalue and outside the main band; the cyan region is where the leading eigenvalue is part of the main band; and the yellow is where the contrast eigenvalue is largest and outside the main band.

The horizontal yellow-blue interface observed in Fig~\ref{fig:heatmap_gen1}(c) for larger $p_{in}^+$ values indicates a direct transition between the prosocial and assortative states as $p_{out}^-$ increases past 0.5. This transition does not involve detectability but rather the shift in the leading eigenvector from homogeneous to contrast that occurs outside the noise band. Equating the contrast and homogeneous eigenvalue expressions, (\ref{eq:lamFpert}) and (\ref{eq:lamHpert}), we find that the transition occurs when $\nu=\mu$, which implies that $\langle A_{out} \rangle =0$ or equivalently $p_{out}^-=0.5$.
%
%%%%%%%%%% tikz %%%%%%%
%%%% first general heatmap %%%
\par
\begin{figure}[!htbp]
\centering
        \begin{tikzpicture}[scale=0.8]
        \node[inner sep=0pt](russell) at (0,0)
        {\includegraphics[width=0.6\linewidth]{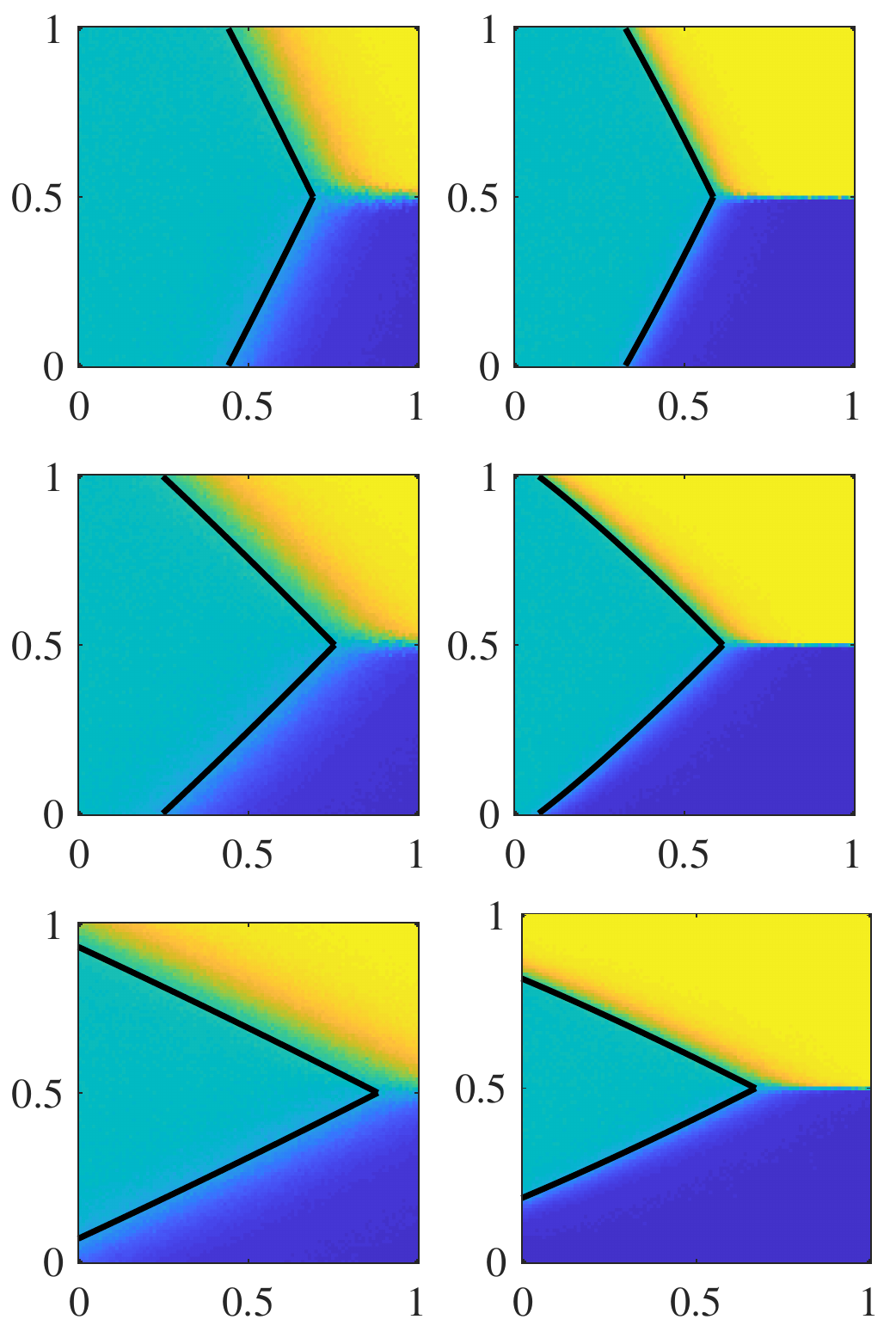}};
        \node (fig2) at (3.7,3.6)
       {\includegraphics[scale=0.168]{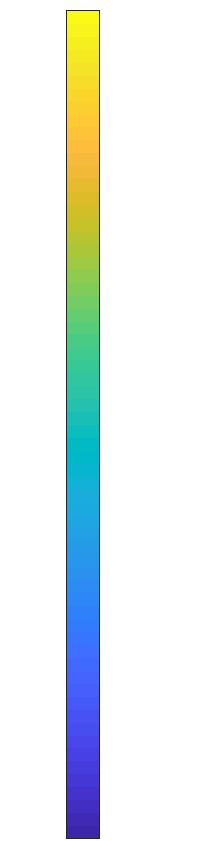}};
       \node at (4.35,4.2)  {$r$};
       \node at (4.35,2.8)  {$h$};
       \node at (3.85,4.8)  {\footnotesize $1$};
       \node at (3.85,3.75)  {\footnotesize $0$};
       \node at (3.85,3.15)  {\small$\frac{1}{2}$};
       \node at (3.85,2.3)  {\footnotesize $1$};
       \draw[line width=0.2mm,|->] (4.1,3.55)--(4.1,5.05);
       \draw[line width=0.2mm,|->] (4.1,3.5)--(4.1,2.1);

        \draw[line width=0.25mm,-] (-4,5.4)--(3.2,5.4);
        \draw[line width=0.25mm,-] (-4,5.4)--(-4,-5);

        \node at (0,6.1)  {$d$};
        \node at (-1.8,5.7)  {\small $0.15$};
        \node at (1.7,5.7)  {\small $0.75$};

        \node at (-4.9,0)  {\large $\frac{d_{in}}{d_{out}}$};
        \node at (-4.3,3.45)  {\small $2$};
        \node at (-4.3,0)  {\small $1$};
        \node at (-4.3,-3.7)  {$\frac{1}{2}$};

        \node at (-2.1,1.8)  {$p_{in}^+$};
        \node at (-3.4,3.0)  {$p_{out}^-$};

        \draw [fill=white,white] (3.5,3.5) rectangle (3.68,3.55);
        \end{tikzpicture}
    \caption{Final state heat maps for increasing network density $d = (d_{in} + d_{out})/2$ and decreasing density ratio $d_{in}/d_{out}$. The assortativity and homogeneity are integrated via the $z$ metric (see Fig.~\ref{fig:heatmap_gen1}).}
 \label{fig:heatmap_gen1_subs}
\end{figure}
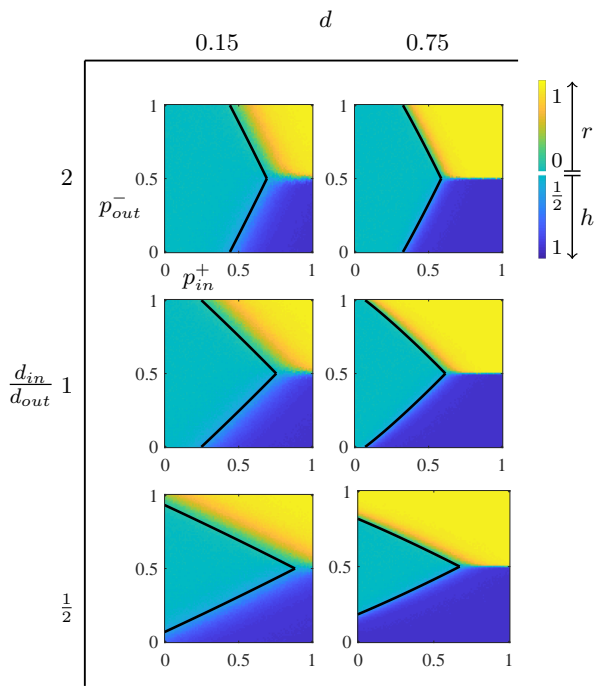

Figure~\ref{fig:heatmap_gen1_subs} shows how the density ratio $d_{in}/d_{out}$ and overall network density $d = (d_{in} + d_{out})/2$ affect the assortative and prosocial transition curves. As $d_{in}/d_{out}$ increases, the transition curves become steeper, implying that denser regions of the connectivity network have more influence on the community structure than sparse regions. 

\section{\label{sec:discussion}Discussion}
This paper has contributed to two distinct areas of signed network research --- community structure and structural balance theory, linking them via the impact of the former upon the latter. First, we summarize and discuss our results on community and, more broadly, network structure, a subject of relevance to both unsigned and signed networks. We then turn to structural balance dynamics, an intrinsically signed network avenue of research. As it is particularly applicable to social systems, we speculate as to connections between our results and the dynamics of conflicts.

\subsection{\label{sec:disc_structure}Structure}

We have elucidated the spectrum of unweighted and undirected signed networks generated by a two-community stochastic block model via two independent methods, perturbation analysis and, in the Appendices, a random matrix theory treatment that extends prior work on unsigned networks \cite{nadakuditi_graph_2012, zhang_spectra_2014}. The expected matrix, $\langle \matr{A} \rangle$, in the block model can be decomposed into two signals --- a homogeneous eigenvector, $\matr{u}_H$, related to the expected tie value, $\mu$, over the network and a contrast eigenvector, $\matr{u}_C$, related to the half-difference, $\nu$, between the expected ingroup and outgroup tie values and which encodes the community structure. These signal eigenvectors exhibit transitions at the points where they merge with the main spectral band associated with the noise produced by the zero-mean random matrix $\matr{X}$. There are four potential transitions corresponding to the intersections of the two signals with the positive and negative edges of the main band. For the contrast eigenvector, these intersections induce the assortative and disassortative transitions respectively and mark changes in community detectability. The homogeneous eigenvector undergoes sociality transitions, prosocial and antisocial, in which emergence from the noise band signifies an overall tendency toward the formation of cooperative or conflictual relationships respectively with other nodes.

We derived analytical expressions for the signal eigenvalues in the presence of the noise by performing a perturbation expansion in which the contributions from the noise $\matr{X}$ were treated as small corrections to the eigenvalues of  $\langle \matr{A} \rangle$. The expressions, (\ref{eq:lamFpert}) and (\ref{eq:lamHpert}), reveal a second-order correction proportional to the average tie variance and are symmetric under the interchange of $\mu$ and $\nu$. The same expressions are derived in the Appendices using random matrix theory along with the formula for the main band edges that is a straightforward modification of Wigner's semicircle law. The transition conditions, (\ref{eq:nucritical}) and (\ref{eq:mucritical}), were determined by equating the signal eigenvalues to the band edge eigenvalues.

We note that our results can be applied to unsigned networks by taking the ingroup and outgroup animosities to be zero so that negative tie formation is prohibited. In this case, our expressions for the homogeneous and contrast eigenvalues reduce to the the sparse-limit forms reported in Refs.~\cite{nadakuditi_graph_2012, zhang_spectra_2014} for the two leading eigenvalues of the adjacency matrix \footnote{Neglecting the $\mu^2$ and $\nu^2$ contributions to $\sigma^2$ in (\ref{eq:sigma2munu}) and inserting into (\ref{eq:lamFpert}) yields $N(d_{in}-d_{out})/2 + (d_{in}+d_{out})/(d_{in}-d_{out})$ for the contrast eigenvalue. Inserting into (\ref{eq:lamHpert}) yields $1+(d_{in}+d_{out})/2$ for the homogeneous eigenvalue}. These forms, however, do not manifest the $\mu$, $\nu$ interchange symmetry, an essential feature of the signed network case.

An important difference between unsigned and signed networks concerns which eigenvectors may represent community structure. In unsigned networks, the assortative community structure can be represented by the second eigenvalue (when past the detectability threshold) but not the first \cite{nadakuditi_graph_2012, zhang_spectra_2014}; the homogeneous and contrast eigenvalues of $\langle \matr{A} \rangle$ grow in proportion to $\mu$ and $\nu$ respectively and the former is always greater than the latter when only positive ties are allowed (unless $\langle A_o \rangle = 0$). Whereas in signed networks, the ordering of $\mu$ and $\nu$ is not so restricted and so the first eigenvalue may signify community structure while the second signifies prosocial structure. Relatedly, while the number of outlying eigenvalues is equal to the number of communities in unsigned networks \cite{ChaGirOtt2009}, this need not be the case in signed networks. For instance, the case of equal ingroup affinity and outgroup animosity treated in Sec.~\ref{sec:regimes} has only one outlying eigenvalue but two communities.

One last remark concerns the sociality transitions. Whereas both signed and unsigned networks can exhibit the prosocial transition, only signed ones can transition to antisocial structure since ties can never be negative in unsigned networks. Typically, community structure is taken to connote the existence of multiple communities as it is linked to the community detection problem and the assignment of nodes to communities. The sociality transitions, which are not relevant to the community assignment problem, involve network structure more generally rather than community structure per se. For a signed network, however, it is meaningful to attempt to discern whether there is an overall tendency, irrespective of ingroup and outgroup distinctions, toward the formation of positive versus negative ties. The existence of such a global tendency therefore provides a sense in which the network can be construed as comprising a single community. The prosocial transition provides a spectral signature for concluding that the network forms a single community in which relationships are generally friendly or cooperative. Conversely, the antisocial transition provides a signature of a single (anti)community marked by hostility, a Hobbesian state of all against all. The intermediate case, where the homogeneous eigenvalue does not appear outside the noise band, can be taken as marking the absence of a single community with a definitive inclination toward positive or negative tie formation.

\subsection{\label{sec:disc_balance}Dynamics}

We investigated structural balance dynamics in the presence of initial community structure generated by the two-identity stochastic block model. These dynamics completely connect all nodes and allow for three broad regimes of final state: an assortative regime in which the two final factions completely align with the two identity blocks; a non-assortative regime in which the two final factions are randomly composed with respect to identity; and a homogeneous regime of a single faction with only positive ties. Since the dominant eigenvector of the initial network drives its structural balance dynamics and determines its final state, our spectral analysis allows us to chart the parameter conditions under which each of these states will emerge. The dynamical ascendance of the leading eigenvector implies that the transitions in which the homogeneous and contrast eigenvalues are just equal to the positive edge of the main spectral band, will correspond to regime boundaries.  The assortative transition marks the boundary between the non-assortative and assortative regimes and the prosocial transition divides the non-assortative and homogeneous regimes. The boundary between the homogeneous and assortative regimes represents a reversal in the ranking of the homogeneous and contrast eigenvalues rather than a transition involving the noise band. The theoretically-predicted boundaries were found to agree with the simulation results obtained by solving the structural balance model over many random initial networks.

Taking networks generated by the stochastic block model and evolving them under the structural balance model entails a dramatic shift in the processes governing network evolution (beyond the change from stochastic to deterministic): a process in which tie formation in a dyad depends only upon its ingroup or outgroup identity is replaced by one where the friendly or hostile relationships among mutual neighbors drives tie value evolution regardless of identity. One rationale for such a shift can be provided by assuming that there is a qualitative change in the nature of the interactions. For instance, hostile relationships characterized by insulting words or gestures may be replaced by physical violence. A second rationale could involve, not a change in the nature of interactions, but a growing awareness that hostile interactions have the potential to become much more prevalent. For example, in a country or region containing two broad identity types, such as ethnicity or ideology, in which individuals or small gangs sporadically clash (either within or across identity lines), a sudden collapse of the central government may lead to a growing sense of looming systemic violence. In either of these rationales, nodes are motivated to seek and maintain allies so that another node's status as the enemy of an enemy or friend of a friend becomes a crucial determinant in relationship formation and evolution.

Our results may inform debates about the interplay of identity and power in conflicts under anarchy consisting of many actors such as insurgencies, civil wars, and international relations. For ethnic conflict, the shift in models discussed above is supported by the observation that the turn from non-violent to violent conflict represents a qualitative change in dynamics \cite{BruLai1998}. In the literature on civil wars and ethnic violence, some theories stress mechanisms in which ethnic or religious identity plays an intrinsic role in producing high levels of polarization and violence along identity lines while other theories stress the role of micro-processes of conflict among local actors rather than a pre-existing identity schism \cite{BruLai1998,KalKoc2007,CedVog2017}. As node identity plays no role in the network evolution, structural balance dynamics is consistent with the latter view. However, the sharp transition to the assortative state shows that the dynamics can lock in initial differences in identity-driven community structure even when they are not large, a behavior consistent with the observation that the polarization and violence in ethnic civil wars often appears to be disproportionate to the initial level of ethnic tension. But the existence of the non-assortative regime implies that identity polarization will not arise when the initial structure is sufficiently weak. Additionally, it has been argued that, contrary to some theories, there is no inherent difference in the dynamics between ideological and ethnic civil wars in terms of their potential for polarization and violence \cite{KalKoc2007}. Our results are consistent with such a claim as it is the initial relationships that matter regardless of whether they are due to similar ethnicity or similar ideology.

Finally, we note a few potential directions for future research for both community structure and structural balance dynamics. As with the unsigned case, spectral analysis of signed networks with community structure could be extended to systems with multiple communities, directed ties, and more realistic network statistics such as non-uniform degree distributions. The structural balance model we used is very simple and, problematically, leads to tie strengths which blow up in finite time. Accordingly, the extent to which the dynamical transitions we have identified persist for more realistic implementations of structural balance dynamics should be explored. More empirical work is also needed to understand the conditions under which real networks can be reasonably modeled by structural balance dynamics or variants thereof.

Code reproducing select results from this paper is available online \footnote{\url{https://github.com/mmtree/Community_signed_networks}}.

\begin{acknowledgments}
This work was funded in part by the Big Data for Genomics and Neuroscience Training Grant under the National Institute of Health Grant No.T32LM012419-03 and by the Office of Naval Research under Grant No. N00014-16-1-2919.
\end{acknowledgments}
\appendix
\section{Spectrum of random matrix \texorpdfstring{$\matr{X}$}{X}}
\label{appendix1}
The signal eigenvalues of $\matr{A}$ are only visible if they are distinguishable from the noise in the system. In this section we find $\mathcal{O}(||\matr{X}||_2)$ as well as $\gamma$ the spectral edge of the noise matrix $\matr{X}$ using random matrix theory.
\par
We start by characterizing the distribution of eigenvalues.
The empirical spectral distribution (e.s.d.) of a random matrix $\matr{X}$ is defined by
\begin{align}
\label{eq:esd}
    \rho(z) = \frac{1}{N} \sum_{i=1}^n \delta(z-\omega_i),
\end{align}
where $\omega_i$ are the eigenvalues of $\matr{X}$.
\par
Wigner's semicircle distribution Eq.~(\ref{eq:WSd}) defines the eigenvalue density function for a symmetric random matrix size $N$ with i.i.d.\ entries having variance $m^2$,
\begin{align}
\label{eq:WSd}
    \phi(z) = \frac{1}{2 \pi N m^2}\sqrt{4Nm^2 -z^2}.
\end{align}
According to the Wigner limit theorem, the spectra of certain symmetric random matrices converge in distribution to Wigner's semicircle distribution \cite{wigner_distribution_1958, bai_spectral_2010}, so that
\begin{align}
    \lim_{N\rightarrow \infty}\int_{-\infty}^c \rho(x)dx = \int_{-\infty}^c \phi(x)dx
\end{align}
Although the semicircle law originally applies only to random symmetric matrices with equal variances for all entries \cite{wigner_distribution_1958,bai_convergence_1988,erdos_universality_2011}, further inquiry has determined that random symmetric block Toeplitz matrices \cite{li_limit_2011,basu_limiting_2012} also weakly converge to the semicircle law under certain conditions.
$\matr{X}$ is a random symmetric block Toeplitz matix with distribution variances $\sigma_{in}^2$ and $\sigma_{out}^2$ in the on and off diagonal blocks respectively and therefore has a Wigner semicircle distribution of eigenvalues for some variance parameter $m^2$ which we have yet to determine. The edges of the seimicircle enclose the band of eigenvalues in the interval $(-2m\sqrt{N}, 2m\sqrt{N})$. As $N \rightarrow \infty$, $\lambda_1(\matr{X}) = 2m\sqrt{N}$.
\par
We find the variance parameter by bounding $||\matr{X}||_2$ with the Frobenius norm \cite{golub_matrix_1996},
\begin{align}
    \frac{1}{\sqrt{N}}||\matr{X}||_F \leq &||\matr{X}||_2 \leq ||\matr{X}||_F,
    \label{eq:X2_bound}
\end{align}
where
\begin{align}
   ||\matr{X}||_F &= \sqrt{\sum_{j=1}^{\frac{N^2}{2}} |X_{in}|^2 + \sum_{j=1}^{\frac{N^2}{2}} |X_{out}|^2}.
   \label{eq:X_frobenius}
\end{align}
The ingroup and outgroup sums are distributed as
\begin{align}   
   \sum_{j=1}^{\frac{N^2}{2}} |X_{in}|^2 &\sim \mathcal{N}\biggl(\frac{N^2}{2}\sigma_{in}^2,\frac{N^2}{2}Var(|X_{in}|^2)\biggl)\\
   \sum_{j=1}^{\frac{N^2}{2}} |X_{out}|^2 &\sim \mathcal{N}\biggl(\frac{N^2}{2}\sigma_{out}^2,\frac{N^2}{2}Var(|X_{out}|^2)\biggl).
\end{align}
The random variable $Z_2$ denoting the inside of the square root in (\ref{eq:X_frobenius}) is therefore distributed as 
\begin{align}
   Z_2 = \sum_{j=1}^{\frac{N^2}{2}} |X_{in}|^2 &+ \sum_{j=1}^{\frac{N^2}{2}} |X_{out}|^2 \sim \mathcal{N}(N^2 \sigma^2, N^2 \xi^2)\\
   \text{where } \xi^2 &= \frac{Var(|X_{in}|^2) + Var(|X_{out}|^2)}{2}.
\end{align}
The expected value of $Z_2$ scales with $\sigma^2 N^2$ while the standard deviation only scales with $\xi N$ meaning $Z_2 = \mathcal{O}(\sigma^2 N^2)$.
Therefore
\begin{align}
   ||\matr{X}||_F &=\mathcal{O}(\sigma N).
\end{align}

Equation~(\ref{eq:X2_bound}) then implies that
\begin{align}
    \sigma \sqrt{N} &\leq ||\matr{X}||_2 \leq \sigma N.
\end{align}
Therefore $||\matr{X}||_2$ must scale with $\sigma$ as
\begin{align}    
    &\mathcal{O}(||\matr{X}||_2) \sim \sigma.
\end{align}
This result combined with the Wigner's semicircle distribution scaling implies that
\begin{align}
  \mathcal{O}(&||\matr{X}||_2) = \sigma \sqrt{N}.
\end{align}
We continue the argument in order to find the leading eigenvalue of $\matr{X}$.
We have found that our variance parameter must scale with $\sigma$, $m = \mathcal{O}(\sigma)$.
\begin{align}
    m = a \sigma = a\sqrt{\frac{\sigma_{in}^2 + \sigma_{out}^2}{2}}\\
    \text{for some constant a} \nonumber
\end{align}
We set $\sigma_{in}^2 = \sigma_{out}^2$ which implies that we have a traditional Wigner matrix with i.i.d.\ entries of variance $\sigma_{in}^2$. This means $m = \sigma_{in} = \sigma$ and therefore $a = 1$. Now let $\sigma_{in}^2 \neq \sigma_{out}^2$ while keeping $\sigma^2$ constant. $m$ scales only with $\sigma$ and therefore must be still equal to $\sigma$. This means that for all variance values, $m = \sigma$.
\par
This implies that the eigenvalue density function of $\matr{X}$ with a diagonal block variance of $\sigma_{in}^2$ and an off diagonal block variance of $\sigma_{out}^2$ is equivalent to the eigenvalue density function of a random matrix with uniform variance $\sigma^2$. This substitution is made in previous derivations of the spectral band of unsigned stochastic block model matrices \cite{nadakuditi_graph_2012}.
\par
Thus, we have determined that the spectra of $\matr{X}$ has a Wigner's semicircle distribution with variance parameter $\sigma$. Therefore the edge of the spectral band $\gamma$ of $\matr{X}$ is
\begin{align}
\label{eq:gamma2}
    \gamma &= 2\sigma\sqrt{N}.
\end{align}
%
%%%%%%%%%%%%%%%%%%%%%%%%%%%%%%%%%%%
%%%%%%%%%%%%%%%%%%%%%%%%%%%%%%%%%%%
\section{Spectra of \texorpdfstring{$\matr{A}$}{A} derived from random matrix theory and complex analysis}
\label{appendix2}
We will now use an alternative method to derive the spectra of $\matr{A}$ using random matrix theory and complex analysis.
In this argument, we use the eigenvalues of the noise matrix $\matr{X}$, whose spectra we have defined in Appendix~\ref{appendix1}, to find the eigenvalues of $\matr{X} + \nu N \matr{u}_C \matr{u}_C^T$. We then take these intermediate eigenvalues and use them to find the eigenvalues of $\matr{A} = \matr{X} + \nu N \matr{u}_C \matr{u}_C^T + \mu N \matr{u}_H \matr{u}_H^T$.
\par
The expected adjacency matrix $\langle \matr{A} \rangle$ is given by Eq.~(\ref{eq:Aavg_clean}). We consider the spectrum obtained by adding just the contribution of the contrast eigenvector, $\nu N \matr{u}_C \matr{u}_C^T$, to the noise matrix which yields the eigenvalue equation
\begin{align}
\label{eqn:add_first_mode}
    (\matr{X} + \nu N \matr{u}_C \matr{u}_C^T) \matr{v} &= z \matr{v}.
\end{align}
We wish to solve for the eigenvalues $z$, and so use the methods of Ref.~\cite{zhang_spectra_2014} to convert Eq.~(\ref{eqn:add_first_mode}) into a trace representation Eq.~(\ref{eqn:trace_format}). We begin by rearranging the terms in Eq.~(\ref{eqn:add_first_mode}) to eliminate the eigenvector $\matr{v}$,
\begin{align}
\label{eq:eliminate_eigvec}
\matr{u}_C^T(z-\matr{X})^{-1}\matr{u}_C &= \frac{1}{\nu N}.
\end{align}
The left hand side of Eq.~(\ref{eq:eliminate_eigvec}) can be written as a sum by performing an eigenvector decomposition on $\matr{X}$,
\begin{align}
\matr{u}_C^T \matr{S} (z\matr{I} - \matr{\Lambda})^{-1}\matr{S}^T \matr{u}_C &= \frac{1}{\nu N}\\
\label{eq:xiuct}
\sum_{i=1}^N \frac{(\matr{x}_i^T \matr{u}_C)^2}{z-\omega_i} &= \frac{1}{\nu N},
\end{align}
where $\matr{S} \matr{\Lambda} \matr{S}^T$ is the eigenvector decomposition of $\matr{X}$ and $\matr{x}_i$ are the eigenvectors of $\matr{X}$ (as well as the columns of $\matr{S}$). We find that $N(\matr{x}_i^T \matr{u}_C)^2 \sim \mathcal{X}_1^2$ and therefore $\mathds{E}[(\matr{x}_i^T \matr{u}_C)^2] = \frac{1}{N}$ and $Var[(\matr{x}_i^T \matr{u}_C)^2] = \frac{2}{N^2}$. This allows us to make the approximation $(\matr{x}_i^T \matr{u}_C)^2 = 1/N$ in Eq.~(\ref{eq:xiuct}) giving us
\begin{align}
    \label{eqn:z_lam_sub}
   \sum_{i=1}^N \frac{1/N}{z-\omega_i} &= \frac{1}{\nu N}\\
    \label{eqn:trace_format}
   \frac{1}{N}Tr (z-\matr{X})^{-1} &= \frac{1}{\nu N}.
\end{align}

We define $f(z)$ as follows,
\begin{align}
\label{eq:f(z)}
     f(z) = Tr (z-\matr{X})^{-1}.
\end{align}
The values of $z$ that satisfy $f(z) = \frac{1}{\nu}$ are the eigenvalues of the matrix $\matr{X}+\nu N \matr{u}_C \matr{u}_C^T$. $f(z)$ has simple poles where $z = \omega_i$, $f(z) \rightarrow -\infty$ as $z \nearrow \omega_i$ and $f(z) \rightarrow \infty$ as $z \searrow \omega_i$. $f(z)$ is a continuous function within the interval $z \in [\omega_i, \omega_{i-1}]$, therefore for each interval $f(z) = \frac{1}{\nu}$ for some value $z \in [\omega_i, \omega_{i-1}]$. This means the eigenvalues $z_i$ and $\omega_i$ are interlaced with the leading eigenvalue $z_1 > \omega_1$.
The largest solution to $f(z) = \frac{1}{\nu}$ Eq.~(\ref{eq:f(z)}) is the leading eigenvalue of $\matr{X} + \mu N \matr{u}_C \matr{u}_C^T$.
\par
We now can repeat this process to find a formula for both leading eigenvalues by adding the homogeneous signal in addition to the contrast signal back into the noise matrix and solving for the resulting eigenvalues $\lambda$ \cite{zhang_spectra_2014}.

The new eigenvalue equation becomes
\begin{align}
\label{eqn:add_second_mode}
    (\matr{X} + \nu N \matr{u}_C \matr{u}_C^T + \mu N \matr{u}_H \matr{u}_H^T) \matr{v} &= \lambda \matr{v}.
\end{align}
We solve this equation for the eigenvalues $\lambda_i$ using the same method used to solve Eq.~(\ref{eqn:add_first_mode}) and which is detailed in Ref.~\cite{zhang_spectra_2014}. The resulting equation $g(\lambda)$ has a similar form to $f(z)$ but with an additional term,

\begin{align}
    \frac{1/N}{\lambda - z_1} + \sum_{i=2}^N \frac{1/N}{\lambda-z_i} = \frac{1}{\mu N}\\
    \label{eqn:approx_g}
    g(\lambda) = \frac{1}{\lambda - z_1} + \sum_{i=2}^N \frac{1}{\lambda-z_i}.
\end{align}
The values of $\lambda$ that satisfy $g(\lambda) =\frac{1}{\mu}$ are the eigenvalues of the matrix $\matr{X} + \nu N \matr{u}_C \matr{u}_C^T + \mu N \matr{u}_H \matr{u}_H^T$. Without loss of generality, assume $\nu > \mu$, meaning we have added the largest eigenvalue mode to the noise matrix followed by the second largest eigenvalue mode.
When $|\lambda - z_1| >> \frac{1}{N}$, $\frac{1}{\lambda-z_1} = \mathcal{O}(1)$ and $\sum_{i=2}^N \frac{1}{\lambda -z_i} = \mathcal{O}(N)$, Because $\frac{1}{\lambda -z_1}$ is the dominant term only when $|\lambda - z_1| = \mathcal{O} (1/N^2)$ and the spectral values $z_i$ that constitute the spectral band of $\matr{X} + \nu N \matr{u}_C \matr{u}_C^T$ are interlaced with the spectrum of $\matr{X}$, we may approximate $g(\lambda)$ with our previous function $f(\lambda) = Tr(\lambda-\matr{X})^{-1}$ for all $\lambda$ values away from $z_1$.  $g(\lambda)$ has a singularity at $z_1$ meaning there is an additional solution to $g(\lambda) = \frac{1}{\mu}$ when $\lambda \approx z_1$.
\par
Therefore the signal eigenvalues, $\lambda_H$ and $\lambda_C$ are the largest magnitude solutions to $f(\lambda) = \frac{1}{\nu}$ and $f(\lambda) = \frac{1}{\mu}$.
We can find an analytical form for $f(\lambda)$ by taking advantage the Stieltjes transform representation of $Tr(\matr{X}-\lambda)^{-1}$.
\par
The Stieltjes transform $S_{\rho}(\lambda)$ of density $\rho(t)$ is a function of the complex variable $\lambda$ and is defined outside the real interval $I$,
\begin{align}
\label{eq:Stiel}
    S_{\rho}(\lambda) = \int_{I} \frac{\rho(t)}{\lambda-t} dt,  \  \ \lambda\in \mathds{C} \backslash  I.
\end{align}
The normalized trace of $(\matr{X}-\lambda)^{-1}$ is equivalent to the Stieltjes transform of the spectral density of $\matr{X}$ \cite{bai_spectral_2010},
\begin{align}
\label{eq:st_eq_of_trace}
    \frac{1}{N}Tr(\matr{X}-\lambda)^{-1} = S_{\rho}(\lambda) = \int_{I}\frac{\rho(x)}{x-\lambda}dx,
\end{align}
where the e.s.d.\ has been previously defined, Eq.~(\ref{eq:esd}).
We may substitute the eigenvalue density function
$\phi(x)$ for the e.s.d.\ integrand in the Stieltjes transform since these functions converge in distribution \cite{bai_spectral_2010},
\begin{align}
\label{eq:st_sub}
    S_{\rho}(\lambda) = \int_{I}\frac{\rho(x)}{x-\lambda}dx = \int_{I}\frac{\phi(x)}{x-\lambda}dx.
\end{align}
Substituting Eq.~(\ref{eq:WSd}) for $\phi(x)$ yields
\begin{align}
    S_{\phi}(\lambda) &= \int_{I}\frac{\phi(x)}{x-\lambda}dx = \frac{1}{2 \pi N \sigma^2}\int_{-2\sqrt{N}\sigma}^{2\sqrt{N}\sigma}\frac{\sqrt{4N\sigma^2-x^2}}{x-\lambda}dx.
\end{align}
We solve this integral using multiple changes of variables. Our argument is adapted from previous work \cite{bai_spectral_2010}. Letting $x = 2\sqrt{N}\sigma \cos(y)$, the above becomes
\begin{align}
     S_{\phi}(\lambda) &= \frac{1}{\pi} \int_{0}^{2 \pi} \frac{\sin^2(y)}{2\sqrt{N}\sigma \cos(y) -\lambda}dy \\
     &= \frac{1}{\pi} \int_0^{2 \pi} \frac{(\frac{e^{iy} - e^{-iy}}{2i})^2}{2\sqrt{N}\sigma (\frac{e^{iy} + e^{-iy}}{2}) - \lambda} dy.
\end{align}
The second change of variables is $\zeta = e^{iy}$ which gives
\begin{align}
\label{eq:S_phi}
     S_{\phi}(\lambda) = \frac{i}{4 \pi \sigma \sqrt{N}} \oint_{|\zeta|=1} &\frac{(\zeta^2-1)^2}{\zeta^2(\zeta^2 -\frac{\lambda}{\sigma \sqrt{N}} \zeta +1)}d\zeta.\\
     \text{Let } h(\zeta) = &\frac{(\zeta^2-1)^2}{\zeta^2(\zeta^2 -\frac{\lambda}{\sigma\sqrt{N}} \zeta +1)}.
\end{align}
The function $h(\zeta)$ has three poles,
\begin{align}
\label{eq:pole0}
    \zeta_0 &= 0\\
\label{eq:pole1}
    \zeta_1 &= \frac{\lambda+\sqrt{\lambda^2 - 4N\sigma^2}}{2\sigma \sqrt{N}}\\
\label{eq:pole2}
    \zeta_2 &= \frac{\lambda - \sqrt{\lambda^2 - 4N \sigma^2}}{2\sigma \sqrt{N}}.
\end{align}
We must determine which poles are inside the radius $|\zeta| = 1$ and then use the residue theorem to compute the integral.
$\zeta_0$ is a pole of order 2 and is inside the contour. Note that $\zeta_1 \zeta_2 = 1$, and therefore if $|\zeta_2| \neq |\zeta_1|$ then only one of these poles can be inside the contour. We find, with the argument to follow, that $\zeta_2$ is the pole inside the contour for $\lambda$ values for which $Im(\lambda) >0$ and $\zeta_1$ is the pole inside the contour for $\lambda$ values for which $Im(\lambda) <0$.
\par
We find that $\zeta_2$ is the enclosed pole for $Im(\lambda)>0$ by first supposing that $Re(\lambda)>0$. It follows that $Re(\sqrt{\lambda^2-4N\sigma^2}) >0$ and $Im(\sqrt{\lambda^2-4N\sigma^2})>0$ and therefore $|\lambda-\sqrt{\lambda^2-4N\sigma^2}|<|\lambda+\sqrt{\lambda^2-4N\sigma^2}|$ which reveals that $|\zeta_2|<|\zeta_1|$. Let us now suppose that $Re(\lambda)<0$. It follows that $Re(\sqrt{\lambda^2-4N\sigma^2}) <0$ and $Im(\sqrt{\lambda^2-4N\sigma^2})>0$, and therefore, by the same argument as above, once again $|\zeta_2|<|\zeta_1|$, meaning that $\zeta_2$ is the enclosed pole.
\par
We use a repetitive argument to show that $\zeta_1$ is the enclosed pole for $Im(\lambda)<0$. Supposing that $Re(\lambda)<0$, we find that $Re(\sqrt{\lambda^2-4N\sigma^2}) >0$ and $Im(\sqrt{\lambda^2-4N\sigma^2}) >0$ which informs us that $|\zeta_1|< |\zeta_2|$. Now supposing that $Re(\lambda)>0$, we find that $Re(\sqrt{\lambda^2-4N\sigma^2}) <0$ and $Im(\sqrt{\lambda^2-4N\sigma^2}) >0$ and therefore $|\zeta_1|< |\zeta_2|$, meaning that $\zeta_1$ is the enclosed pole.
\par
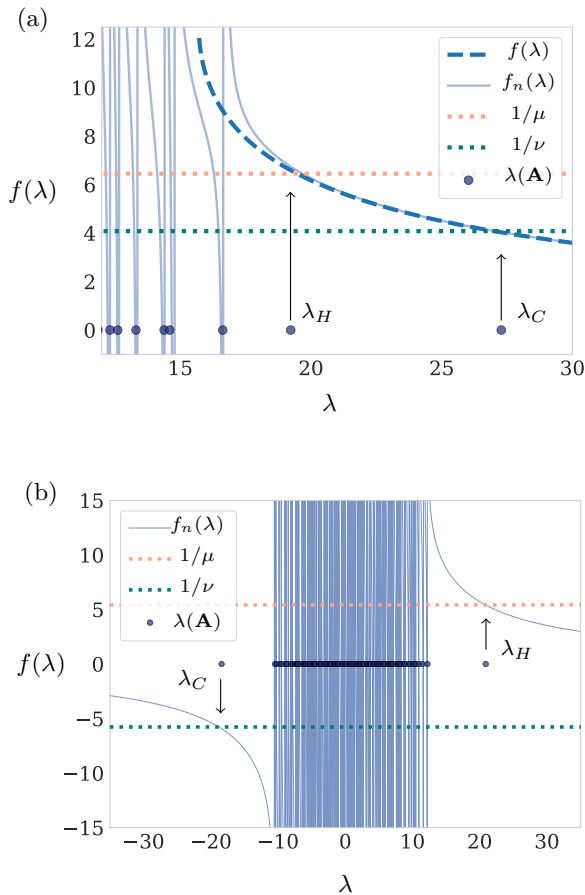
\begin{figure}[!htbp]
\centering
\subfloat{%
\resizebox{\linewidth}{!}{
\begin{tikzpicture}[scale=1]
\node[inner sep=0pt](russell) at (0,0)
 {\includegraphics[width=0.9\linewidth]{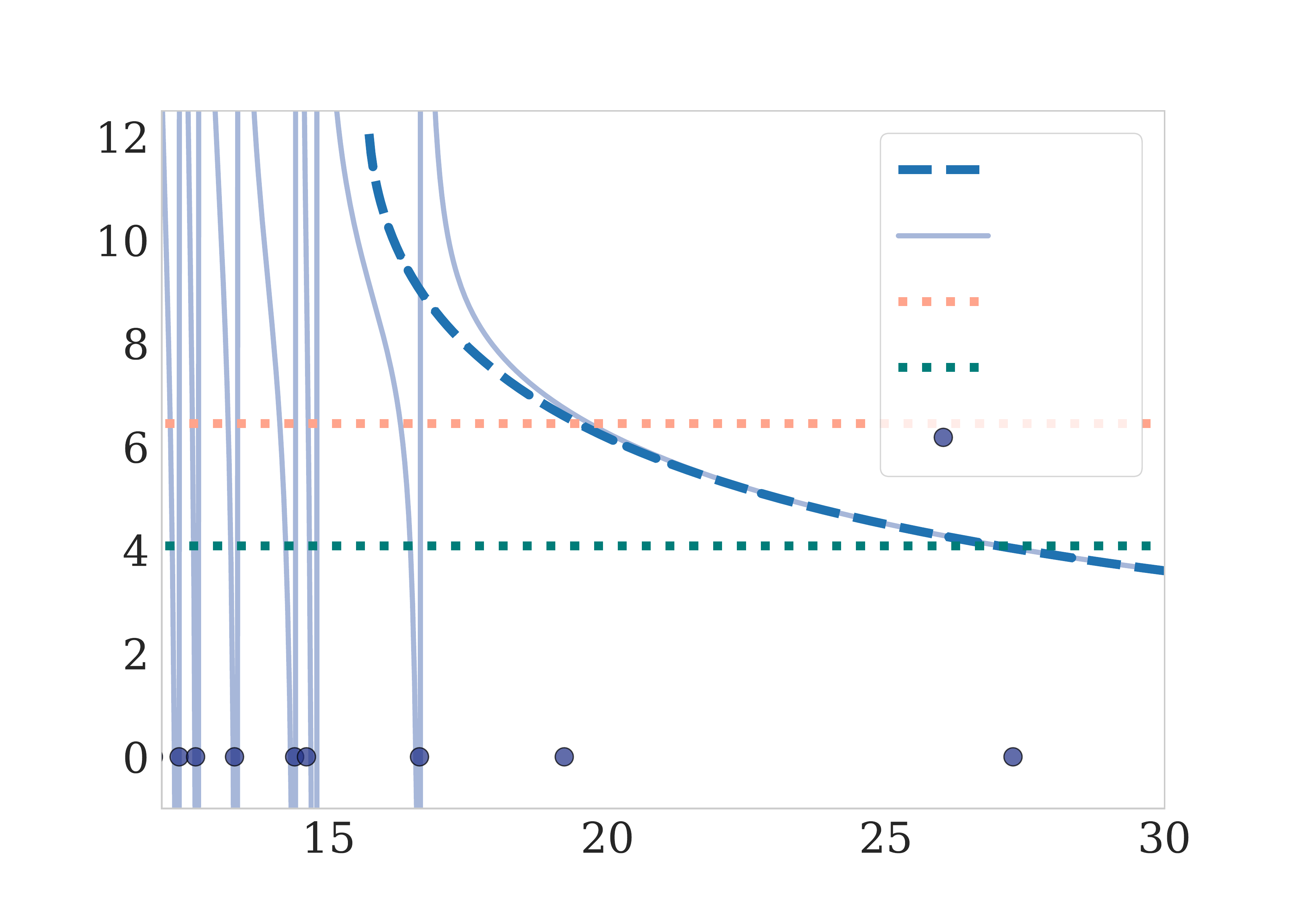}};

 \node at (-3.8,2.2) {(a)};

 \draw[->](2.2,-1.65)--(2.2,-0.8);
 \draw[->](-0.5,-1.65)--(-0.5,0);

 \node at (-3.8,0) {$f(\lambda)$};
 \node at (0,-2.7) {$\lambda$};
 \node at (2.6,-1.55) {\footnotesize $\lambda_C$};
 \node at (-0.15,-1.55) {\footnotesize $\lambda_H$};

 \node at (2.55,1.8) {\scriptsize $f(\lambda)$};
 \node at (2.55,1.4) {\scriptsize $f_n(\lambda)$};
 \node at (2.55,1) {\scriptsize $1/\mu$};
 \node at (2.55,0.6) {\scriptsize $1/\nu$};
 \node at (2.55,0.2) {\scriptsize $\lambda(\matr{A})$};
\end{tikzpicture}
}
}\vfil
\subfloat{
\resizebox{\linewidth}{!}{
\begin{tikzpicture}[scale=1]
\node[inner sep=0pt](russell) at (0,0)
 {\includegraphics[width=0.9\linewidth]{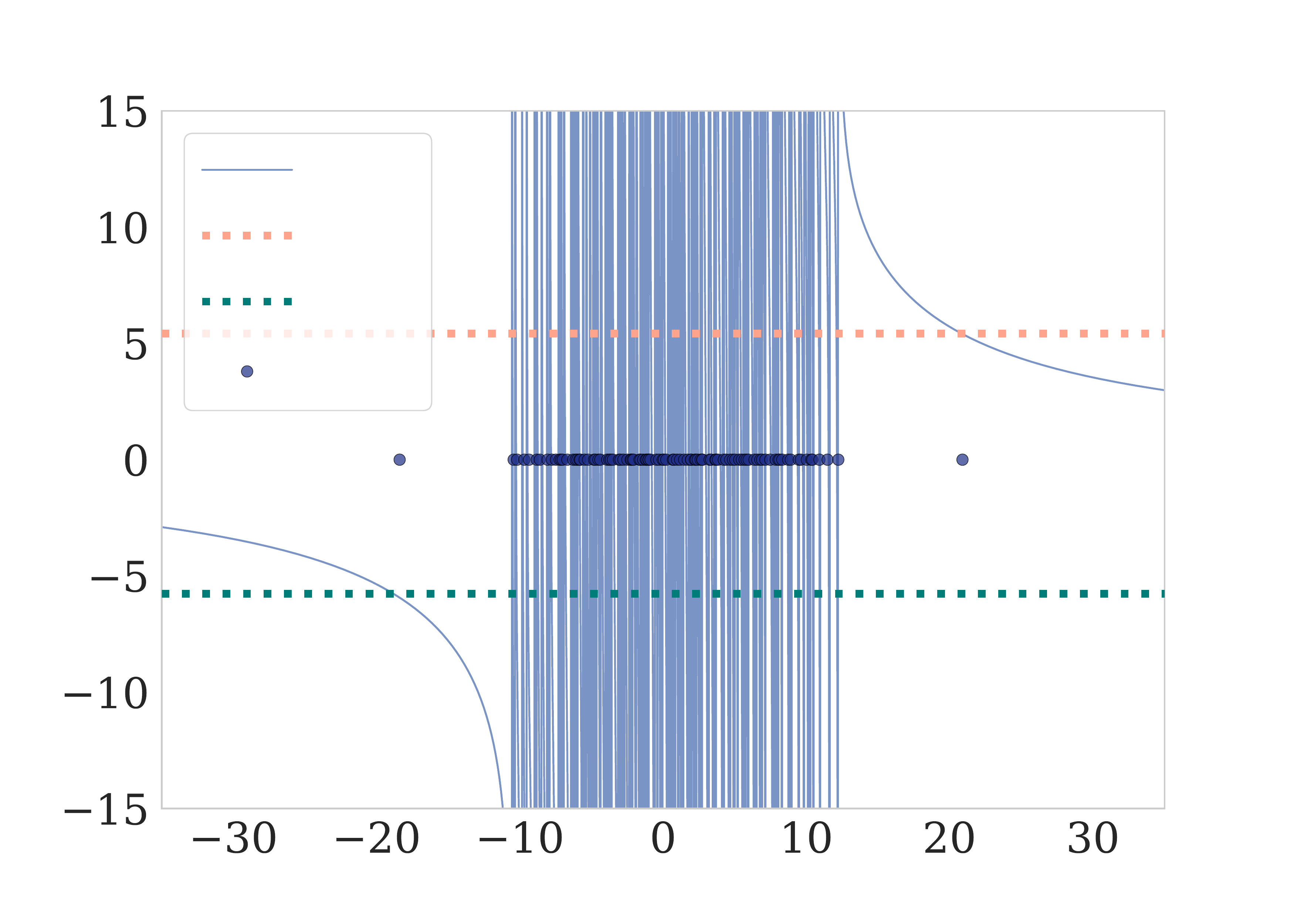}};
  \node at (-3.8,2.2) {(b)};

 \draw[->](1.9,0.2)--(1.9,0.6);
 \draw[->](-1.5,-0.18)--(-1.5,-0.62);

 \node at (2.3,0.2) {\footnotesize $\lambda_H$};
 \node at (-1.85,-0.18) {\footnotesize $\lambda_C$};

 \node at (-1.8,1.8) {\scriptsize $f_n(\lambda)$};
 \node at (-1.8,1.4) {\scriptsize $1/\mu$};
 \node at (-1.8,1) {\scriptsize $1/\nu$};
 \node at (-1.8,0.55) {\scriptsize $\lambda(\matr{A})$};

 \node at (0.1,-2.8) {$\lambda$};
 \node at (-3.8,0) {$f(\lambda)$};
 \end{tikzpicture}
 }
}
\caption{Intersections of the function $f(\lambda)$ and $1/\mu$ and $1/\nu$ yield eigenvalue locations. (a) $f_n(\lambda)$ is the numerical solution to $f(\lambda)$ and intersects with $1/\mu$ and $1/\nu$ at $\lambda_H$ and $\lambda_C$.  $f(\lambda)$ is only defined away from the spectral band; the analytical solution diverges from the numerical approximation upon approaching the spectral edge. (b) The case where $\lambda_H$ and $\lambda_C$ are on opposite sides of the spectral band as the signal eigenvalues can be either positive or negative in signed networks.}
\label{fig:ped_complex}
\end{figure}

We now know that $\zeta_1$ is inside the contour and $\zeta_2$ is outside for $Im(\lambda)<0$ while $\zeta_2$ is inside the contour and $\zeta_2$ is outside for $Im(\lambda)>0$. We use the Residue theorem and the three poles Eqs.~(\ref{eq:pole0}), (\ref{eq:pole1}), and (\ref{eq:pole2}) to finish solving Eq.~(\ref{eq:S_phi}) by integrating,
\begin{align}
 \oint_{|\zeta|=1} h(\zeta)d\zeta &= 2 \pi i [Res(h,\zeta_0) + Res(h,\zeta_*)]\\
 \text{where } \zeta_* &= \left\{
      \begin{array}{@{}ll@{}}
        \zeta_1 & ,\text{ for } Im(\lambda) <0\\
        \zeta_2 & ,\text{ for } Im(\lambda)>0.
      \end{array}\right.
\end{align}
Evaluating the residues yields
\begin{align}
  \oint_{|\zeta|=1} h(\zeta)d\zeta =2 \pi i \bigg[\frac{\lambda}{\sigma \sqrt{N}} \pm \frac{\sqrt{\lambda^2-4N\sigma^2}}{\sigma \sqrt{N}} \bigg].
 \end{align}
 The $\pm$ in the above expression results from the two different residues for $Im(\lambda)>0$ and $Im(\lambda)<0$. We now multiply our result by the constant term to finish solving Eq.~(\ref{eq:S_phi}),
 \begin{align}
 S_{\phi}(\lambda) &= \frac{-\lambda \pm \sqrt{\lambda^2 - 4N\sigma^2}}{2N\sigma^2}.
 \end{align}
Now we have a closed form solution for the Stieltjes transform of the spectral density of $\matr{X}$ which gives us an analytical formula for $f(\lambda)$ in the region when $\lambda$ is outside of the spectral band, $|\lambda|> 2 \sigma \sqrt{N} = \gamma$,
\begin{align}
 \label{eq:tr_analytic}
 f(\lambda) = Tr(\lambda-\matr{X})^{-1} &= \frac{\lambda \pm \sqrt{\lambda^2 - 4N\sigma^2}}{2\sigma^2}, \ \ \ \ |\lambda|>\gamma.
\end{align}
We now solve for the eigenvalue $\lambda$ of $\matr{A}$ that corresponds to the contrast signal,
\begin{align}
  f(\lambda) &= \frac{1}{\nu}\\
  \frac{\lambda \pm \sqrt{\lambda^2 - 4N \sigma^2}}{2\sigma^2} &= \frac{1}{\nu}\\
  \label{eq:z_F_complex}
  \implies \lambda = \nu N + \frac{\sigma^2}{\nu}&,\ \ |\nu| \geq \frac{\sigma}{\sqrt{N}}.
\end{align}
If $|\nu| \geq \frac{\sigma}{\sqrt{N}}$ then $|\lambda| \geq |\gamma|$, and the contrast eigenvalue $\lambda$ is outside the spectral band, otherwise the leading eigenvalue is included in the spectral band. The analogous formula for the eigenvalue corresponding to the homogeneous signal is found by replacing $\nu$ by $\mu$ in Eq.~(\ref{eq:z_F_complex}). Figure~\ref{fig:ped_complex} shows how the intersections between $f(\lambda)$ and $1/\mu$ and $1/\nu$ generate the eigenvalue locations. We have therefore found the same solutions for the signal eigenvalues of $\matr{A}$ using random matrix theory and complex analysis as done via perturbation theory in the main text, Eqs.~(\ref{eq:lamFpert}) and  (\ref{eq:lamHpert}).
%%%%%%%%%%%%%%%%%%%%%%%%%%%%
%%%%%%% tikz pic %%%%%%%%%%%
%%%%%%%%%%%%%%%%%%%%%%%%%%%%

\bibliographystyle{apsrev4-2}
%\bibliography{MyLibrary,MGRefs}% Produces the bibliography via BibTeX.

\end{document}